\documentclass[usenatbib,usegraphicx]{mn2e}
\usepackage{float}
\usepackage{color}
\usepackage[fleqn]{amsmath}
\usepackage[colorlinks=true, citecolor=blue]{hyperref}
\usepackage{amssymb}
\newcommand{\be}{\begin{equation}}
\newcommand{\ee}{\end{equation}}
\definecolor{red}{rgb}{1,0.0,0.0}

\definecolor{darkgreen}{rgb}{0.0,0.5,0.0}

\newcommand{\beq}{\begin{eqnarray}}  
\newcommand{\eeq}{\end{eqnarray}}  
 
\newcommand{\apj}{ApJ}  
\newcommand{\apjs}{ApJS}  
\newcommand{\apjl}{ApJL}  
\newcommand{\aj}{AJ}  
\newcommand{\mnras}{MNRAS}  
  
\newcommand{\aap}{A\&A}  
  
\newcommand{\araa}{ARA\&A}  
\newcommand{\nat}{Nature}

\newcommand{\ly}{{\ifmmode{{\rm Ly}\alpha}\else{Ly$\alpha$}\fi}}
\newcommand{\hMpc}{{\ifmmode{h^{-1}{\rm Mpc}}\else{$h^{-1}$Mpc }\fi}}  
\newcommand{\hGpc}{{\ifmmode{h^{-1}{\rm Gpc}}\else{$h^{-1}$Gpc }\fi}}  
\newcommand{\hmpc}{{\ifmmode{h^{-1}{\rm Mpc}}\else{$h^{-1}$Mpc }\fi}}  
\newcommand{\hkpc}{{\ifmmode{h^{-1}{\rm kpc}}\else{$h^{-1}$kpc }\fi}}  
\newcommand{\hMsun}{{\ifmmode{h^{-1}{\rm {M_{\odot}}}}\else{$h^{-1}{\rm{M_{\odot}}}$}\fi}}  
\newcommand{\hmsun}{{\ifmmode{h^{-1}{\rm {M_{\odot}}}}\else{$h^{-1}{\rm{M_{\odot}}}$}\fi}}  
\newcommand{\Msun}{{\ifmmode{{\rm {M_{\odot}}}}\else{${\rm{M_{\odot}}}$}\fi}}  
\newcommand{\msun}{{\ifmmode{{\rm {M_{\odot}}}}\else{${\rm{M_{\odot}}}$}\fi}}

\newcommand{\rand}{{\ifmmode{{\mathcal{R}}}\else{${\mathcal{R}}$ }\fi}}

\newcommand{\sub}[1]{\mbox{\scriptsize{#1}}}

\newcommand{\bds}[1]{\boldsymbol{ #1 }}

\newcommand{\eq}[2]{\begin{equation} \label{eq:#1} #2 \end{equation}}

\def\aj{AJ}%
\def\araa{ARA\&A}%
\def\apj{ApJ}%
\def\apjl{ApJ}%
\def\apjs{ApJS}%
\def\aap{A\&A}%
\def\aapr{A\&A~Rev.}%
\def\mnras{MNRAS}%
\def\prd{Phys.~Rev.~D}%
\def\prl{Phys.~Rev.~Lett.}%
\def\nat{Nature}%
\title[Spin evolution and feedback of supermassive black holes]
{Spin evolution and feedback of supermassive black holes in cosmological simulations}

\author[S.~Bustamante and V.~Springel]
{\parbox{17cm}
{Sebastian Bustamante$^{1,2}$ and Volker~Springel$^{3}$}\vspace*{0.1cm}\\
$^1$Heidelberger Institut f\"{u}r Theoretische Studien,
  Schloss-Wolfsbrunnenweg 35, 69118 Heidelberg, Germany\\
$^2$Zentrum f\"ur Astronomie der Universit\"at Heidelberg, Astronomisches Recheninstitut, M\"{o}nchhofstr. 12-14, 69120 Heidelberg, Germany\\
$^3$Max-Planck-Institut f\"ur Astrophysik, Karl-Schwarzschild-Str. 1, D-85748, Garching, Germany }

\begin{document}
\pagerange{\pageref{firstpage}--\pageref{lastpage}}
\pubyear{2018}

\maketitle

\label{firstpage}

\begin{abstract}
It is well established that the properties of supermassive black holes (BH) and their host galaxies are correlated through scaling relations. While hydrodynamical cosmological simulations have begun to account for the co-evolution of BHs and galaxies, they typically have neglected the BH spin, even though it may play an important role in modulating the growth and feedback of BHs. Here we introduce a new sub-grid model for the BH spin evolution in the moving-mesh code {\small AREPO} in order to improve the physical faithfulness of the BH modelling in galaxy formation simulations. We account for several different channels of spin evolution, in particular gas accretion through a Shakura-Sunyaev $\alpha$-disc, chaotic accretion, and BH mergers.  For BH feedback, we extend the IllustrisTNG model, which considers two different BH feedback modes, a thermal quasar mode for high accretion states and a kinetic mode for low Eddington ratios, with a self-consistent accounting of spin-dependent radiative efficiencies and thus feedback strength. We find that BHs with mass $M_{\sub{bh}}\lesssim 10^{8}\, {\rm M}_{\odot}$ reach high spin values as they typically evolve in the coherent gas accretion regime, in which consecutive accretion episodes are aligned. On the other hand, BHs with mass $M_{\sub{bh}}\gtrsim 10^{8}\, {\rm M}_{\odot}$ have lower spins as BH mergers become more frequent, and their accretion discs fragment due to self-gravity, inducing chaotic accretion. We also explore the hypothesis that the transition between the quasar and kinetic feedback modes is mediated by the accretion mode of the BH disc itself, i.e.~the kinetic feedback mode is activated when the disc enters the self-gravity regime instead of by an ad-hoc switch tied to the BH mass. We find excellent agreement between the galaxy and BH populations for this approach and the fiducial TNG model with no spin evolution. Furthermore, our new approach  alleviates a tension in the galaxy morphology -- colour relation of the original TNG model.
\end{abstract}

\begin{keywords}
Black hole physics -- quasars: supermassive black holes -- accretion, accretion discs -- methods: numerical
\end{keywords}

%==================================================================================================
\section{Introduction}
%==================================================================================================

It is remarkable finding that most, quite possibly all, galaxies host a supermassive black hole (BH) at their centres \citep[e.g.][]{Greene2010}. The origin of these black holes (i.e.~the ``seeds'' that subsequently grew) in the high redshift universe is still unclear, but it appears relatively certain that much of the mass of the black hole population was accumulated through gas accretion \citep{Soltan1982}. The accretion processes themselves are accompanied by a substantial release of energy, which is most dramatically seen in the emission of powerful quasars \citep{LyndenBell1969}.
 
The relatively tight scaling relations between galaxy properties and the masses of the central supermassive BH \citep{Magorrian1998, Ferrarese2000, Haering2004} suggest that the BHs are not merely spectators in the galaxy formation process, but rather influence it in decisive ways. This view has motivated the incorporation of black hole evolution and feedback processes in modern theories of galaxy formation \citep{Kauffmann2000}.  In particular, many simulation models explain the quenching transition of galaxies from the star forming blue cloud to the red sequence of elliptical galaxies as a result of BH feedback \citep{DiMatteo2005, Springel2005, Sijacki2007}. In large clusters of galaxies, the BHs can furthermore offset radiative cooling losses through sporadically going through phases of active galactic nuclei (AGN) activity, thereby preventing excessive cooling flows. Another importance of this AGN feedback lies in its ability to expel baryons even from deep potential wells of galaxy groups and clusters, helping to understand the observed trends of the baryon content in these objects \citep{McCarthy2007, Puchwein2008}.

The recent discovery of gravitational waves from stellar-mass black holes \citep{Abbott2016} has additionally fuelled the interest in the merger processes of supermassive black holes, which are inevitable events during hierarchical structure formation. In the future, coalescing BH binaries in the relevant mass range should become detectable by pulsar timing arrays and the Laser Interferometer Antenna (LISA). These unprecedented experiments will shed light on the BH assembly process, but they also require theoretical predictions for the expected merger rates and a characterisation of the precursors of BH binaries. The BH mergers will emit copious amounts of gravitational waves, leading to a recoil of the merger remnant. As a result, they may be even ejected from the host galaxy in certain situations. In any case, the  timescale of return of the remnant BH to the centre of the galaxy is important as during this phase the regulation of a quasi-static hot atmosphere in the halo may temporarily interrupted, allowing star formation to resume and potentially impact the evolution of massive galaxies.

Supermassive black holes are therefore a key element of any modern theory of galaxy formation and evolution, and they are now routinely incorporated both in semi-analytic \citep[e.g.][]{Croton2006, Henriques2015} and hydrodynamical simulation models \citep[e.g.][]{Vogelsberger2014, Schaye2015}.  Despite this central importance of supermassive BHs, the physical processes governing their evolution, especially with respect to the gas accretion and the associated feedback processes, are only poorly understood. The modelling in cosmological hydrodynamic simulations is hence very sketchy, and typically encapsulated in heuristic sub-grid models. Commonly, black holes are simply treated as sink particles of a given mass, and all modelling of the physics neglects the one other important parameter characterising the black holes, namely their angular momentum. This is a significant limitation, because the BH spin affects the radiative efficiency of accretion processes decisively. In fact, while it is believed that typically of order 5\% of the accreted rest mass energy of a non-spinning BH is released as radiation, this can go up to about 40\% for highly spinning black holes \citep{Blandford1977, Tchekhovskoy2011}. In addition, the spin of merging BHs has a strong influence on the gravitational wave emission, and thus also on the expected recoil velocity of the merger remnant \citep{Campanelli2007}.

For all these reasons, modelling BH spins appears to be a promising extension of current treatments of black hole physics. Corresponding simulations would then not only yield predictions for the BH mass function and its evolution over time, for the lifetimes and activity patterns of quasars, their clustering, and for their overall influence on the galaxy formation process, they would also be able to make predictions for the statistics of BH spin, including for the typical spin configurations expected in supermassive BH mergers. This in turn would inform attempts to forecast gravitational wave signals and their frequency, and elucidate the importance of spin for the energy release through gas accretion. It is therefore timely to go beyond the first generation of BH models and implement more detailed physical models for BH spin in the current cosmological codes for galaxy formation.

At first sight, it may seem conceptually simple to relate the BH growth to the specific angular momentum content of the accreted gas, but the huge range of scales that are in reality still unresolved in cosmological simulations can not simply be ignored easily. In fact, the accretion flows that actually develop close in to the BH can give rise to angular momentum transport, in addition to an important coupling between a preexisting BH spin and the surrounding gas, causing phenomena such as Lense-Thirring precession. This necessitates the construction of explicit sub-grid models for the BH spin evolution which are guided by analytic theory, and which respond to the larger-scale boundary conditions set by the cosmological simulation. A few such models have already been constructed. \citet{Fanidakis2011} pioneered such treatments in the context of semi-analytic modelling of galaxy formation. \citet{Dubois2014c} study the BH spin evolution in hydrodynamical cosmological simulations in a post-processing fashion. Recently, and independently from our work, \citet{Fiacconi2018} presented a first effort using the moving-mesh code {\small AREPO}, which we also employ here. Our model is similar in spirit to theirs, but differs in a number of details. Importantly, we consider the back-reaction on the feedback efficiency and couple the spins to the black hole feedback mode. In particular, we conjecture to relate the transition in feedback modes, introduced in the IllustrisTNG black hole treatment in an ad-hoc fashion by \citet{Weinberger2017}, to the spin evolution, and test the impact this has.

This paper is structured as follows. In Section~\ref{SecBHSpins}, we begin by reviewing the basic problem of how BH spin evolves through gas accretion and BH binary coalescence. In Section~\ref{SecBHmodel}, we detail our black hole growth and feedback model, as well as its coupling to the spin evolution. We then discuss our numerical implementation of these processes and our simulation set in Section~\ref{SecNumerics}. Our results for the statistics of BH spin in cosmological simulations are presented in Section~\ref{SecResultsStatistics}, while results for the impact of our spin modelling through a novel feedback switch on the Section~\ref{SecBHSpinFeedback}. Finally, we summarise our findings and give our conclusions in Section~\ref{SecConclusions}.

%==================================================================================================
\section{Spin evolution of Black Holes} \label{SecBHSpins}
%==================================================================================================

In this section, we present a model for the evolution of the BH spin. Besides its mass, the spin of a BH is an important property as it affects the efficiency with which a BH can process parts of the matter in an accreted disc into radiation; it impacts the strength and emission direction of radio jets; it possibly regulates the radio-loudness of the host galaxy; and it determines both the detailed gravitational wave emission signal and the distribution of recoil velocities in BH-BH mergers, which is important for forecasting these events and the subsequent evolution of the BH remnant. 

The spin is simply defined as the BH angular momentum with respect to the centre-of-mass frame, and is commonly expressed as:
%eeeeeeeeeeeeeeeeeeeeeeeeeeeeeeeeeeeeeeeeeeeeeeeeeeeeeeeeeeeeeeeeeeeeeeeeeeeeee
\eq{SpinBH}
{ J_{\sub{bh}} = |a|\frac{GM_{\sub{bh}}^2}{c}, }
%eeeeeeeeeeeeeeeeeeeeeeeeeeeeeeeeeeeeeeeeeeeeeeeeeeeeeeeeeeeeeeeeeeeeeeeeeeeeee
where $a$ is the dimensionless spin parameter ($0\leq |a| \leq 1$), $M_{\sub{bh}}$ the black hole mass, $G$ the gravitational constant and $c$ the vacuum speed of light. This parametrisation results from assuming a Kerr metric for rotating bodies, in which the maximum allowed spin is $GM_{\sub{bh}}^2/c$. In what follows, we assume two main channels through which BHs can be spun up or down, namely gas accretion and BH mergers.

%--------------------------------------------------------------------------------------------------
\subsection{Gas accretion}
%--------------------------------------------------------------------------------------------------

BHs in galactic centres process matter into radiation through a surrounding accretion disc that is fed by cold gas inflows from galactic scales. If a BH is accreting above $1\%$ of its Eddington rate,  we assume that a classical thin, optically thick Shakura-Sunyaev $\alpha$-disc \citep{Shakura1973} settles around it \citep{Fanidakis2011}. On the other hand, for accretion rates below this value, the $\alpha$-disc model is no longer appropriate and advection dominated disc models (ADAF) would have to be considered. However, given that in this case the gas does not carry enough angular momentum to modify the spin of the BH significantly, we simply turn off the gas channel under these conditions.

Once an accretion disc is formed, the orbiting gas loses its angular momentum through viscous torques caused by magnetic fields \citep{LyndenBell1969} and radiates away the binding gravitational energy, moving thus inwards and reaching the radius of the innermost stable circular orbit (isco) of the BH. After this point, further angular momentum and energy losses are no longer required as the gas will then be accreted on a very short timescale. 

The radius of the isco can be expressed in terms of the BH spin as \citep{Bardeen1972}:
%eeeeeeeeeeeeeeeeeeeeeeeeeeeeeeeeeeeeeeeeeeeeeeeeeeeeeeeeeeeeeeeeeeeeeeeeeeeeee
\eq{isco}
{ \hat{r}_{\sub{isco}} = \frac{r_{\sub{isco}}}{R_g} = 3 + Z_2 \pm [ (3-Z_1)(3+Z_1+2Z_2) ]^{1/2}, }
%eeeeeeeeeeeeeeeeeeeeeeeeeeeeeeeeeeeeeeeeeeeeeeeeeeeeeeeeeeeeeeeeeeeeeeeeeeeeee
where the negative sign is taken if the disc is counter-rotating with respect to the BH spin (i.e. $a<0$). The positive sign corresponds to the case of co-rotation. For the sake of simplicity, a normalisation with the gravitational radius $R_g$ is used here, which is defined as half of the Schwarzschild radius:
%eeeeeeeeeeeeeeeeeeeeeeeeeeeeeeeeeeeeeeeeeeeeeeeeeeeeeeeeeeeeeeeeeeeeeeeeeeeeee
\eq{Rgrav}
{ R_{\sub{g}} = \frac{R_{\sub{Schw}}}{2} = \frac{GM_{\sub{bh}}}{c}. }
%eeeeeeeeeeeeeeeeeeeeeeeeeeeeeeeeeeeeeeeeeeeeeeeeeeeeeeeeeeeeeeeeeeeeeeeeeeeeee

The quantities $Z_1$ and $Z_2$ depend on the BH spin, and are defined as:
%eeeeeeeeeeeeeeeeeeeeeeeeeeeeeeeeeeeeeeeeeeeeeeeeeeeeeeeeeeeeeeeeeeeeeeeeeeeeee
\begin{eqnarray}
Z_1 &\equiv& 1 + (1-a^2)^{1/3}\left[ (1+a)^{1/3} + (1-a)^{1/3} \right], \\
Z_2 &\equiv& ( 3a^2 + Z_1^2 )^{1/2}.
\end{eqnarray}
%eeeeeeeeeeeeeeeeeeeeeeeeeeeeeeeeeeeeeeeeeeeeeeeeeeeeeeeeeeeeeeeeeeeeeeeeeeeeee

Once a parcel of gas of mass ${\rm d}M_0$ reaches the isco, it will be eventually accreted and its energy and angular momentum per unit mass, $\tilde{e}_{\sub{isco}}$ and $\tilde{l}_{\sub{isco}}$,  will add to the total BH mass and angular momentum, respectively:
%eeeeeeeeeeeeeeeeeeeeeeeeeeeeeeeeeeeeeeeeeeeeeeeeeeeeeeeeeeeeeeeeeeeeeeeeeeeeee
\eq{AccretedMandS}
{ {\rm d}M_{\sub{bh}} = \frac{\tilde{e}_{\sub{isco}}}{c^2}{\rm d}M_0,\ \ \ {\rm d}J_{\sub{bh}} = \tilde{l}_{\sub{isco}}\,{\rm d}M_0. }
%eeeeeeeeeeeeeeeeeeeeeeeeeeeeeeeeeeeeeeeeeeeeeeeeeeeeeeeeeeeeeeeeeeeeeeeeeeeeee

Combining these expressions with equation~(\ref{eq:SpinBH}) yields the following differential equation:
%eeeeeeeeeeeeeeeeeeeeeeeeeeeeeeeeeeeeeeeeeeeeeeeeeeeeeeeeeeeeeeeeeeeeeeeeeeeeee
\eq{DSpinEvolution}
{ \frac{{\rm d}a}{{\rm d} \ln M_{\sub{bh}} } = \frac{1}{M_{\sub{bh}}}\frac{c^3}{G}\frac{\tilde{l}_{\sub{isco}}}{\tilde{e}_{\sub{isco}}} - 2a. }
%eeeeeeeeeeeeeeeeeeeeeeeeeeeeeeeeeeeeeeeeeeeeeeeeeeeeeeeeeeeeeeeeeeeeeeeeeeeeee

Integrating this equation, \citet{Bardeen1970} obtained a solution for the evolution of the spin magnitude:
%eeeeeeeeeeeeeeeeeeeeeeeeeeeeeeeeeeeeeeeeeeeeeeeeeeeeeeeeeeeeeeeeeeeeeeeeeeeeee
\eq{SpinEvol}
{ a^f = \frac{1}{3}\hat{r}_{\sub{isco}}^{1/2}\frac{M_{\sub{bh}}}{M^f_{\sub{bh}}}\left[ 4 - \left\{ 3\hat{r}_{\sub{isco}}\left(\frac{M_{\sub{bh}}}{M^f_{\sub{bh}}}\right)^2 - 2 \right\}^{1/2} \right], }
%eeeeeeeeeeeeeeeeeeeeeeeeeeeeeeeeeeeeeeeeeeeeeeeeeeeeeeeeeeeeeeeeeeeeeeeeeeeeee
where $a^f$ and $M_{\sub{bh}}^f$ correspond to the final spin and BH mass after an accretion episode of mass 
%eeeeeeeeeeeeeeeeeeeeeeeeeeeeeeeeeeeeeeeeeeeeeeeeeeeeeeeeeeeeeeeeeeeeeeeeeeeeee
\eq{AccretedMass}
{M_{\sub{d}} = \frac{M^f_{\sub{bh}} - M_{\sub{bh}}}{1-\epsilon_r}.}
%eeeeeeeeeeeeeeeeeeeeeeeeeeeeeeeeeeeeeeeeeeeeeeeeeeeeeeeeeeeeeeeeeeeeeeeeeeeeee

Here, $\epsilon_r$ is the radiative efficiency of the accretion disc, which measures the fraction of gravitational binding energy released by gas spiralling in towards the BH. In the case of a Kerr BH, the radiative efficiency depends on the spin as follows \citep{NovikovThorne1973}:
%eeeeeeeeeeeeeeeeeeeeeeeeeeeeeeeeeeeeeeeeeeeeeeeeeeeeeeeeeeeeeeeeeeeeeeeeeeeeee
\eq{Epsilon_R}
{ \epsilon_r \equiv 1 - \sqrt{ 1 - \frac{2}{3}\frac{1}{\hat{r}_{\sub{isco}}(a)} }. }
%eeeeeeeeeeeeeeeeeeeeeeeeeeeeeeeeeeeeeeeeeeeeeeeeeeeeeeeeeeeeeeeeeeeeeeeeeeeeee

Note that equation~(\ref{eq:SpinEvol}) can be used to update the BH spin only if $M^f_{\sub{bh}}/M_{\sub{bh}} \leq \hat{r}_{\sub{isco}}^{1/2}$. In the case of $M^f_{\sub{bh}}/M_{\sub{bh}} > \hat{r}_{\sub{isco}}^{1/2}$, a single accretion episode is able to spin up the BH to its maximum value $a=1$; however, we cap the maximum spin to $a = 0.998$ to account for the absorption of photons with angular momentum opposite to the BH spin \citep{Thorne1974}.

\subsubsection{Accreting gas in misaligned discs}

The previous analysis led us to a formula to compute the evolution of the BH spin once a mass $M_d$ has fallen in towards the accretion disc. However, this analysis provides no information about the accretion process and on how much mass is ultimately accreted, i.e.~we do not know the value of $M_d$. In order to estimate this, we assume a general case in which an accretion disc settles around the BH and lies initially misaligned with respect to the BH equatorial plane. Under this configuration, the orbiting gas experiences a torque caused by the Lense-Thirring effect of the spinning BH, which is given by:
%eeeeeeeeeeeeeeeeeeeeeeeeeeeeeeeeeeeeeeeeeeeeeeeeeeeeeeeeeeeeeeeeeeeeeeeeeeeeee
\eq{LenseThirringTorque}
{ \frac{\partial \bds{L}}{\partial t} = \bds{\omega}_{\sub{prec}}\times \bds{L}, }
%eeeeeeeeeeeeeeeeeeeeeeeeeeeeeeeeeeeeeeeeeeeeeeeeeeeeeeeeeeeeeeeeeeeeeeeeeeeeee
where $\bds{\omega}_{\sub{prec}} = (2G/c)\bds{J}_{\sub{bh}}/R^3$ is the precession rate, $\bds{L}$ is the angular momentum per unit area of the disc and $R$ the distance from the BH \citep{Pringle1992}. This torque causes the gas orbits to become unstable and evolve into lower energy states, which in turn makes the plane of the disc precess around the BH spin axis at a frequency ${\omega}_{\sub{prec}}$. If the viscosity of the gas is sufficiently high, the innermost part of the disc aligns with the equatorial plane of the spinning BH, giving rise to a warped disc.

In the $\alpha$-disc solution, the kinematic viscosity of the gas is split into two components, $\nu_1$ and $\nu_2$, which act along velocity gradients parallel and perpendicular to the disc plane, respectively. This allows us to introduce two characteristic timescales: an accretion timescale $t_{\sub{acc}} \equiv R^2/\nu_1$, during which gas drifts inwards and angular momentum is transported outwards through radial viscous dissipation; and a warp timescale $t_{\sub{warp}} \equiv R^2/\nu_2$, during which any vertical disturbance propagates across the disc, e.g.~the warp. 

Introducing the precession timescale $t_{\sub{prec}} \equiv 2\pi/\omega_{\sub{prec}}$, it is possible to estimate the size of the warped region by imposing the condition $t_{\sub{prec}}(R) \lesssim t_{\sub{warp}}(R)$, i.e.~if restoring viscous forces are overcome by frame-dragging torques the disc cannot retain its original shape. Using the following parametrisation of the radial viscosity $\nu_1 = \alpha H^2 \Omega_k$, where $H$ is the disc semi-thickness and $\Omega_k$ the Kepler frequency, the following expression for the warp radius $R_{\sub{warp}}$ is obtained \citep{Volonteri2007}: 
%eeeeeeeeeeeeeeeeeeeeeeeeeeeeeeeeeeeeeeeeeeeeeeeeeeeeeeeeeeeeeeeeeeeeeeeeeeeeee
\eq{Rwarp}
{ \frac{R_{\sub{warp}}}{R_{\sub{Schw}}} = 3.6 \times 10^3 a^{5/8} \left( \frac{M_{\sub{bh}}}{10^8 {\rm M}_{\odot}} \right)^{1/8} f^{-1/4}\left( \frac{\nu_2}{\nu_1} \right)^{-5/8}\alpha^{-1/2}, }
%eeeeeeeeeeeeeeeeeeeeeeeeeeeeeeeeeeeeeeeeeeeeeeeeeeeeeeeeeeeeeeeeeeeeeeeeeeeeee
where $f \equiv L/L_{\sub{Edd}}$ is the Eddington ratio, $L=\epsilon_r \dot{M} c^2$ and $L_{\sub{Edd}} = 4\pi G M_{\sub{bh}}m_p/\kappa$ are the disc and Eddington luminosities, respectively, $\dot{M}$ is the total accretion rate of the system, $m_p$ the proton mass, and $\kappa \approx 0.3\ \mbox{cm}^2\mbox{g}^{-1}$ denotes the electron scattering opacity. 

The warp radius is an important quantity as it provides a characteristic length scale of the accretion disc. Besides, only the material inside it can effectively transfer angular momentum to the BH \citep{Volonteri2007}. The mass of the warped region can be readily estimated as $M_{\sub{warp}} = \dot{M}\,  t_{\sub{acc}}(R_{\sub{warp}})$, where:
%eeeeeeeeeeeeeeeeeeeeeeeeeeeeeeeeeeeeeeeeeeeeeeeeeeeeeeeeeeeeeeeeeeeeeeeeeeeeee
\begin{eqnarray}
\nonumber
t_{\sub{acc}} = \frac{R^2_{\sub{warp}}}{\nu_1} = 3\times 10^6a^{7/8} \left( \frac{M_{\sub{bh}}}{10^8 {\rm M}_{\odot}} \right)^{11/8} \\ 
\label{eq:Tacc}
\times \lambda^{-3/4} \left( \frac{\nu_2}{\nu_1} \right)^{-7/8} \alpha^{-3/2}\ \mbox{yr}.
\end{eqnarray}
%eeeeeeeeeeeeeeeeeeeeeeeeeeeeeeeeeeeeeeeeeeeeeeeeeeeeeeeeeeeeeeeeeeeeeeeeeeeeee

One of the main uncertainties of the $\alpha$-disc solution is the relation between the two viscosities $\nu_1$ and $\nu_2$, which ultimately determines which process, warp alignment or mass accretion, occurs in a shorter timescale. In the thin disc regime, i.e.~$H/R < \alpha \ll 1$, the condition $\nu_2/\nu_1 \approx 1/\alpha^2$ has to be fulfilled \citep{Papaloizou1983}. We use the relation $\nu_2/\nu_1 = 2(1+7\alpha)/(\alpha^2(4+\alpha^2))$ proposed by \citet{Ogilvie1999} that satisfies this condition. Under these assumptions, we obtain that $t_{\sub{warp}}<t_{\sub{acc}}$, which means that the innermost disc aligns with the equatorial plane of the BH quicker than the timescale of mass and angular momentum accretion. This allows us to use $M_{\sub{warp}}$ as a proxy of the accreted mass of a single accretion episode, i.e.~$M_{\sub{d}}\approx M_{\sub{warp}}$. Finally, using equation~(\ref{eq:AccretedMass}), we calculate the final BH mass $M_{\sub{bh}}^f$ needed for evolving the BH spin through equation~(\ref{eq:SpinEvol}).

%.........................................................................
%	Figure 1
%.........................................................................
%Carton of the accretion disc in both regimes
\begin{figure*}
\centering
\includegraphics[width=0.9\textwidth]
{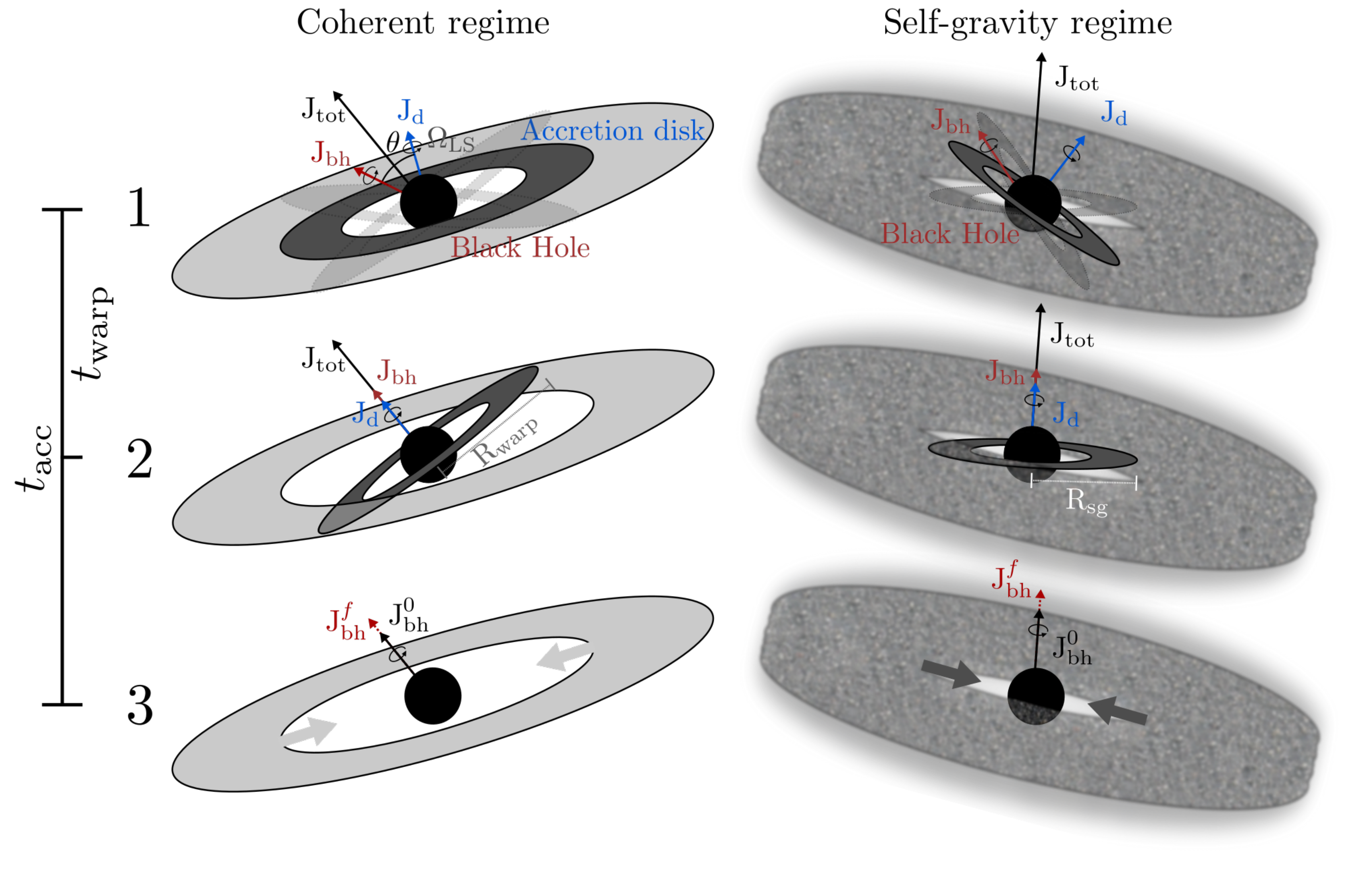}
\caption{\small{Schematic illustration of the spin evolution model. The evolution steps in the coherent (\textit{left}) and self-gravity (\textit{right}) regimes are shown. \textit{Step 1} corresponds to the beginning of the accretion episode, when the innermost disc is precessing about the BH spin due to the Lense-Thirring effect. In the coherent regime there is an outer disc, whereas in the self-gravity regime the outer gas is fragmented and not settled in a disc. In \textit{step 2}, the Lense-Thirring torque eventually aligns/anti-aligns the innermost disc and aligns the BH spin in the direction of the total angular momentum, giving rise to a warped disc. This occurs on the warp timescale $t_{\sub{warp}}$. In the coherent regime, the warped disc has a radius $R_{\sub{warp}}$. In the self-gravity regime the radius is smaller, and is given by $R_{\sub{sg}}$. In \textit{step 3} the warped disc has been already consumed by the BH, and a new accretion episode is already setting in. Given that the accretion timescale is longer, i.e.~$t_{\sub{warp}}<t_{\sub{acc}}$, this is the last step of the accretion episode. In the coherent regime, the new accretion episode inherits the orientation of the outer disc. In the chaotic regime, the direction of the innermost disc angular momentum follows a von Mises distribution function.}}

\label{fig:AccretionDisc}
\end{figure*}
%.........................................................................

\subsubsection{Alignment/Counter-alignment condition}

We have previously concluded that before the content of the innermost disc is accreted, it settles on the equatorial plane of the BH. Additionally, \citet{King2005} established that the warped region, once settled, can align or anti-aligned with respect to the BH spin depending on the initial configuration of the disc angular momentum and the BH spin. In order to highlight the importance of determining the sense of rotation of the gas at the isco, we analyse equation~(\ref{eq:isco}). For a disc counter-rotating with respect to a maximum spinning BH, the radius of the isco is $\hat{r}_{\sub{isco}} = 9$. In the case of co-rotation, $\hat{r}_{\sub{isco}} = 1$. This means that counter-rotating gas needs to lose less energy and angular momentum than co-rotating gas before being ultimately accreted and, therefore, adds up more to the BH mass and spin. On the other hand, co-rotating gas also increases the energy output of the BH, favouring feedback processes.

In what follows, we inspect what condition determines alignment/anti-alignment. We denote the angular momentum of the innermost warped disc as $\bds{J}_{d}$. The total angular momentum is then given by:
%eeeeeeeeeeeeeeeeeeeeeeeeeeeeeeeeeeeeeeeeeeeeeeeeeeeeeeeeeeeeeeeeeeeeeeeeeeeeee
\eq{TotalJ}
{ \bds J_{\sub{tot}} = \bds J_{\sub{bh}} + \bds J_{\sub{d}}. }
%eeeeeeeeeeeeeeeeeeeeeeeeeeeeeeeeeeeeeeeeeeeeeeeeeeeeeeeeeeeeeeeeeeeeeeeeeeeeee

This vector is constant in magnitude and orientation during the whole accretion episode due to conservation of angular momentum in the BH-disc system. In Figure~\ref{fig:AccretionDisc} we illustrate the relative orientations of the vectors for different phases of an accretion episode. \citet{King2005} derived a set of equations that describes the evolution of $\bds J_{\sub{bh}}$ and $\bds J_{\sub{d}}$ during the precession phase:
%eeeeeeeeeeeeeeeeeeeeeeeeeeeeeeeeeeeeeeeeeeeeeeeeeeeeeeeeeeeeeeeeeeeeeeeeeeeeee
\eq{PrecessionEqns}
{ \frac{\rm d}{{\rm d}t}J_{\sub{bh}} ^2 = 0,\ \ \ \frac{\rm d}{{\rm d}t}J_{\sub{d}}\leq 0,\ \ \ \frac{\rm d}{{\rm d}t}\cos\theta_t \geq 0, }
%eeeeeeeeeeeeeeeeeeeeeeeeeeeeeeeeeeeeeeeeeeeeeeeeeeeeeeeeeeeeeeeeeeeeeeeeeeeeee
where $\theta_t$ is the angle subtended between $\bds{J}_{\sub{tot}}$ and $\bds{J}_{\sub{bh}}$. This shows that during this phase, the magnitude of the BH spin remains constant, while the disc angular momentum is continuously decreasing. The direction of the BH spin eventually aligns with the total angular momentum, while the disc will either counter-align or align depending on whether or not the following condition is satisfied \citep{King2005}:
%eeeeeeeeeeeeeeeeeeeeeeeeeeeeeeeeeeeeeeeeeeeeeeeeeeeeeeeeeeeeeeeeeeeeeeeeeeeeee
\eq{AlignmentCond}
{ \cos \theta < -\frac{J_{\sub{d}}}{2J_{\sub{bh}}}, }
%eeeeeeeeeeeeeeeeeeeeeeeeeeeeeeeeeeeeeeeeeeeeeeeeeeeeeeeeeeeeeeeeeeeeeeeeeeeeee
with $\theta$ the angle subtended between $\bds{J}_{\sub{bh}}$ and $\bds{J}_{\sub{d}}$. 

Considering that the angular momentum of the innermost disc is advected inwards, passing trough $R_{\sub{warp}}$, its magnitude can readily be computed as \citep{King2008}:
%eeeeeeeeeeeeeeeeeeeeeeeeeeeeeeeeeeeeeeeeeeeeeeeeeeeeeeeeeeeeeeeeeeeeeeeeeeeeee
\eq{AMWarpedDisc}
{J_{\sub{d}} \approx M_d \Omega_k(R_{\sub{warp}})R_{\sub{warp}}^2 = M_d ( GM_{\sub{bh}}R_{\sub{warp}} )^{1/2}.}
%eeeeeeeeeeeeeeeeeeeeeeeeeeeeeeeeeeeeeeeeeeeeeeeeeeeeeeeeeeeeeeeeeeeeeeeeeeeeee
We have assumed that the gas has circularised on its way to the centre. This leads to the following expression for the ratio of disc to BH angular momentum:
%eeeeeeeeeeeeeeeeeeeeeeeeeeeeeeeeeeeeeeeeeeeeeeeeeeeeeeeeeeeeeeeeeeeeeeeeeeeeee
\eq{RatioAM}
{\frac{J_{\sub{d}}}{2J_{\sub{bh}}} \approx \frac{M_{\sub{d}}}{aM_{\sub{bh}}}\left( \frac{R_{\sub{warp}}}{R_{\sub{Schw}}}\right)^{1/2}. }
%eeeeeeeeeeeeeeeeeeeeeeeeeeeeeeeeeeeeeeeeeeeeeeeeeeeeeeeeeeeeeeeeeeeeeeeeeeeeee

By evaluating this expression at the beginning of every accretion episode, we can determine whether alignment or counter-alignment occurs before the disc is accreted. 

The last aspect to be determined is the orientation of $\bds{J}_{\sub{d}}$. To do so, we assume that it coincides with that of the angular momentum of the neighbouring gas that ultimately feeds the BH-disc system. Although there is a large gap between the physical scale of the gas reservoir in the galactic centre and the scale of the accretion disc, it is reasonable to assume that during the feeding process, when the gas is on its way towards the disc, its angular momentum orientation is preserved.

\subsubsection{Self-gravitating discs}

When the mass accretion rate of the system is sufficiently large, the accretion disc becomes very massive and the effects of self-gravity are no longer negligible. This makes the disc become unstable and fragment into small gas clumps. In order to assess this more quantitatively, we use  Toomre's stability criterion which balances rotational support of the disc against its own gravity:
%eeeeeeeeeeeeeeeeeeeeeeeeeeeeeeeeeeeeeeeeeeeeeeeeeeeeeeeeeeeeeeeeeeeeeeeeeeeeee
\eq{Toomre}
{ Q \equiv \frac{c_s \Omega_k}{\pi G \Sigma_{\sub{d}}}, }
%eeeeeeeeeeeeeeeeeeeeeeeeeeeeeeeeeeeeeeeeeeeeeeeeeeeeeeeeeeeeeeeeeeeeeeeeeeeeee
where $Q$ is the Toomre parameter and $c_s$ is the speed of sound in the gas. For $Q\leq1$ the disc is unstable, which leads to the definition of a characteristic self-gravity radius $R_{\sub{sg}}$ at which $Q(R_{\sub{sg}}) = 1$. For the $\alpha$-disc solution, the following expression is derived \citep{Fanidakis2011}:
%eeeeeeeeeeeeeeeeeeeeeeeeeeeeeeeeeeeeeeeeeeeeeeeeeeeeeeeeeeeeeeeeeeeeeeeeeeeeee
\eq{Rselfgrav}
{ \frac{R_{\sub{sg}}}{R_{\sub{Schw}}} = 1.5 \times 10^3 \epsilon ^{8/27} \left( \frac{M_{\sub{bh}}}{10^8\ \mbox{M}_{\odot}} \right)^{-26/27} f^{-8/27}\alpha^{14/27}. }
%eeeeeeeeeeeeeeeeeeeeeeeeeeeeeeeeeeeeeeeeeeeeeeeeeeeeeeeeeeeeeeeeeeeeeeeeeeeeee

As the parameter $Q$ is a monotonically decreasing function of radius, the disc outside $R_{\sub{sg}}$ is subject to fragmentation whereas the inner disc remains stable. Nevertheless, as only the material within the warp radius can effectively transfer its angular momentum to the BH \citep{Volonteri2007}, we include self-gravity effects only when $R_{\sub{sg}}\leq R_{\sub{warp}}$ \citep{Dubois2014b}. This is the defining condition of what we call the \textit{self-gravity regime}. As opposed to this, we define the \textit{coherent regime} when self-gravity is negligible. In Figure~\ref{fig:AccretionDisc} we illustrate both regimes and the evolution steps in every accretion episode.

Note that during the self-gravity regime, the innermost disc still aligns with the equatorial plane of the BH and its mass is now given by \citep{Fanidakis2011}:
%eeeeeeeeeeeeeeeeeeeeeeeeeeeeeeeeeeeeeeeeeeeeeeeeeeeeeeeeeeeeeeeeeeeeeeeeeeeeee
\eq{Mselfgrav}
{ M_{\sub{sg}} = 2.13 \times 10^5 \epsilon ^{-5/27} \left( \frac{M_{\sub{bh}}}{10^8\ \mbox{M}_{\odot}} \right)^{23/27} f^{5/27}\alpha^{-2/27}\ \mbox{M}_{\odot}. }
%eeeeeeeeeeeeeeeeeeeeeeeeeeeeeeeeeeeeeeeeeeeeeeeeeeeeeeeeeeeeeeeeeeeeeeeeeeeeee
Using $M_{\sub{sg}}$ and $R_{\sub{sg}}$ instead of $M_{\sub{d}}$ and $R_{\sub{warp}}$ in equations (\ref{eq:AccretedMass}) and (\ref{eq:RatioAM}), we can still evolve the magnitude of the BH spin.

Due to the intrinsic complexity of the self-gravity regime, the orientation of the disc angular momentum, or equivalently the material feeding the disc, has been a matter of debate. For instance, \citet{King2008} argued that low cooling times of the gas beyond the self-gravity radius provides an ideal scenario for star formation, which can occur in about a dynamical timescale. This would have disturbing effects on the fragmented disc that most likely will induce random motion of the clumps of gas. \citet{King2008} assumed that the direction of the angular momentum of the gas clumps feeding the inner disc follows an isotropic distribution over $4\pi$ steradians. This corresponds to what is dubbed \textit{chaotic accretion}.

On the other hand, \citet{Volonteri2007} proposed the so-called \textit{coherent accretion}, in which the angular momentum of consecutive accretion episodes is aligned with one another. In the same fashion, \citet{Dubois2014b} always align, in every accretion episode, the disc angular momentum with that of the surrounding gas. We follow this approach only in the coherent regime.

\citet{Dotti2013} assumed a somewhat intermediate approach, in which they assigned two probabilities, $F$ and $1-F$, for an accretion event to have its angular momentum pointing towards the northern or the southern hemisphere as defined by the galactic reference frame, respectively. Following this line of reasoning, we assume that the angular momentum imparted by the gas accreted from large scales is not completely erased. Nevertheless, considering that self-gravity is indeed a stochastic and complex process, the disc angular momentum should not be exactly aligned with the surrounding gas either. This is why we propose a novel approach in which the degree of anisotropy of the fuelling process can be varied, allowing its effects to be studied. In order to do so, we use the von Mises distribution function, a circular analogous of the normal distribution:
%eeeeeeeeeeeeeeeeeeeeeeeeeeeeeeeeeeeeeeeeeeeeeeeeeeeeeeeeeeeeeeeeeeeeeeeeeeeeee
\eq{vonMises}
{ f(\theta|k) = \frac{e^{k\cos{\theta}}}{2\pi I_0(k)}, }
%eeeeeeeeeeeeeeeeeeeeeeeeeeeeeeeeeeeeeeeeeeeeeeeeeeeeeeeeeeeeeeeeeeeeeeeeeeeeee
where $\theta$ is the polar angle, $k$ is the concentration parameter that quantifies the anisotropy degree and $I_0(k)$ gives the Bessel function of zeroth-order. The $z$-axis is still defined by the direction of the angular momentum of the surrounding gas. 

\begin{figure}
\centering
\includegraphics[trim={0cm 0cm 0cm 0cm},width=0.47\textwidth]
{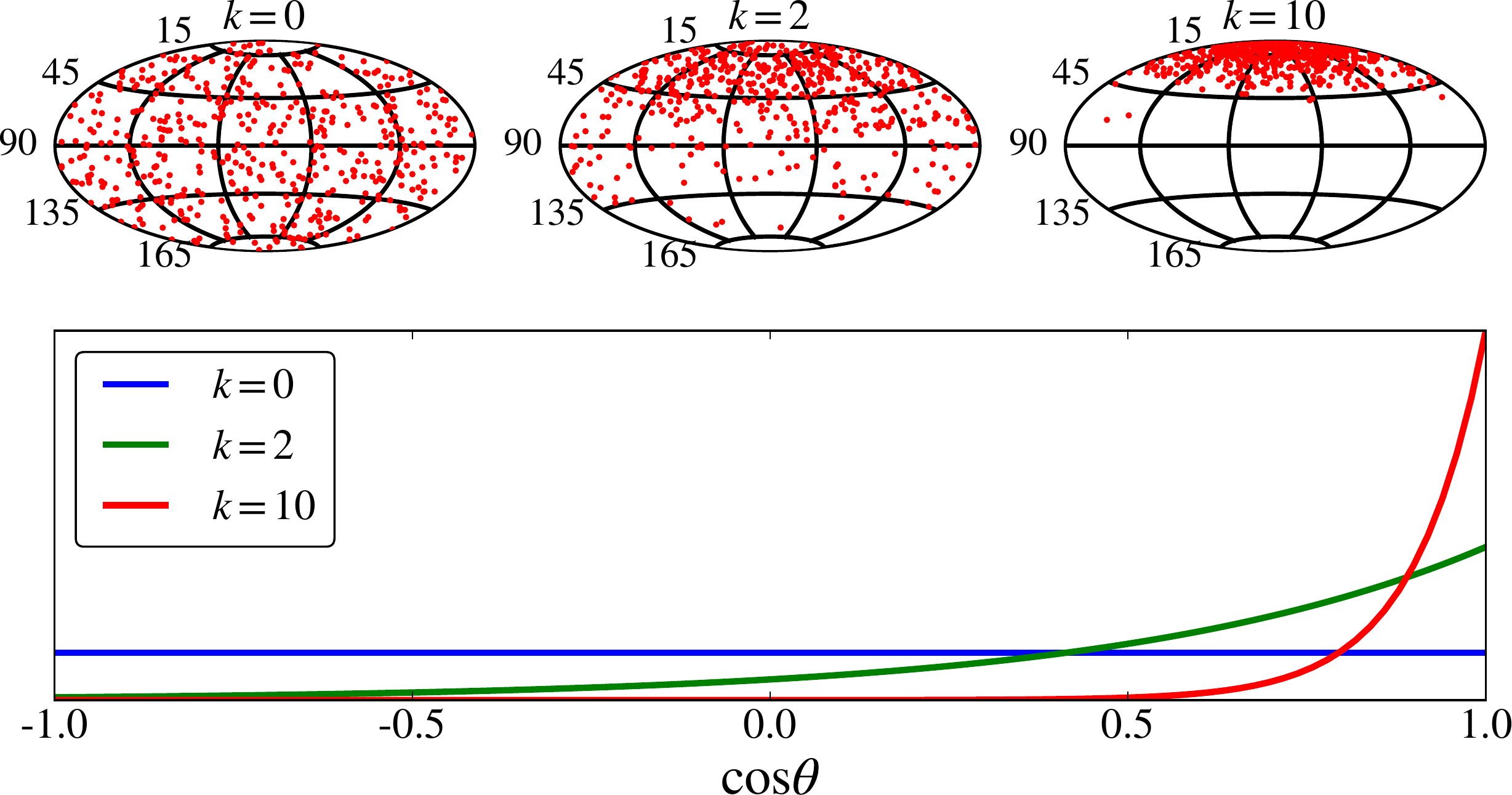}
\caption{\small{\textit{Top panels:} Sky projections of the von Mises distribution for three different anisotropy degrees, ranging from complete isotropy ($k=0$) to a more coherent distribution ($k=10$). \textit{Bottom panel:} PDF of the polar angle for the three choices of the anisotropy degree parameter illustrated in the upper panels.}}

\label{fig:vonMises}
\end{figure}

In Figure~\ref{fig:vonMises}, we show different parameter choices for $k$. For instance, $k=0$ corresponds to an isotropic distribution that recovers chaotic accretion. For $k=10$, the distribution is very concentrated within the cone $\theta < \pi/4$ around the direction of the surrounding gas. For $k$ sufficiently large, we recover the coherent accretion as well.

%.........................................................................
%	Figure 2
%.........................................................................
%Anisotropy in fuelling process
%.........................................................................

%--------------------------------------------------------------------------------------------------
\subsection{Coalescence of BH binaries}
%--------------------------------------------------------------------------------------------------

%.........................................................................
%	Figure 3
%.........................................................................
%Reference system respect which gravitational recoil velocity is applied
\begin{figure}
\centering
\includegraphics[trim={0cm 0cm 0cm 0cm},width=0.3\textwidth]
{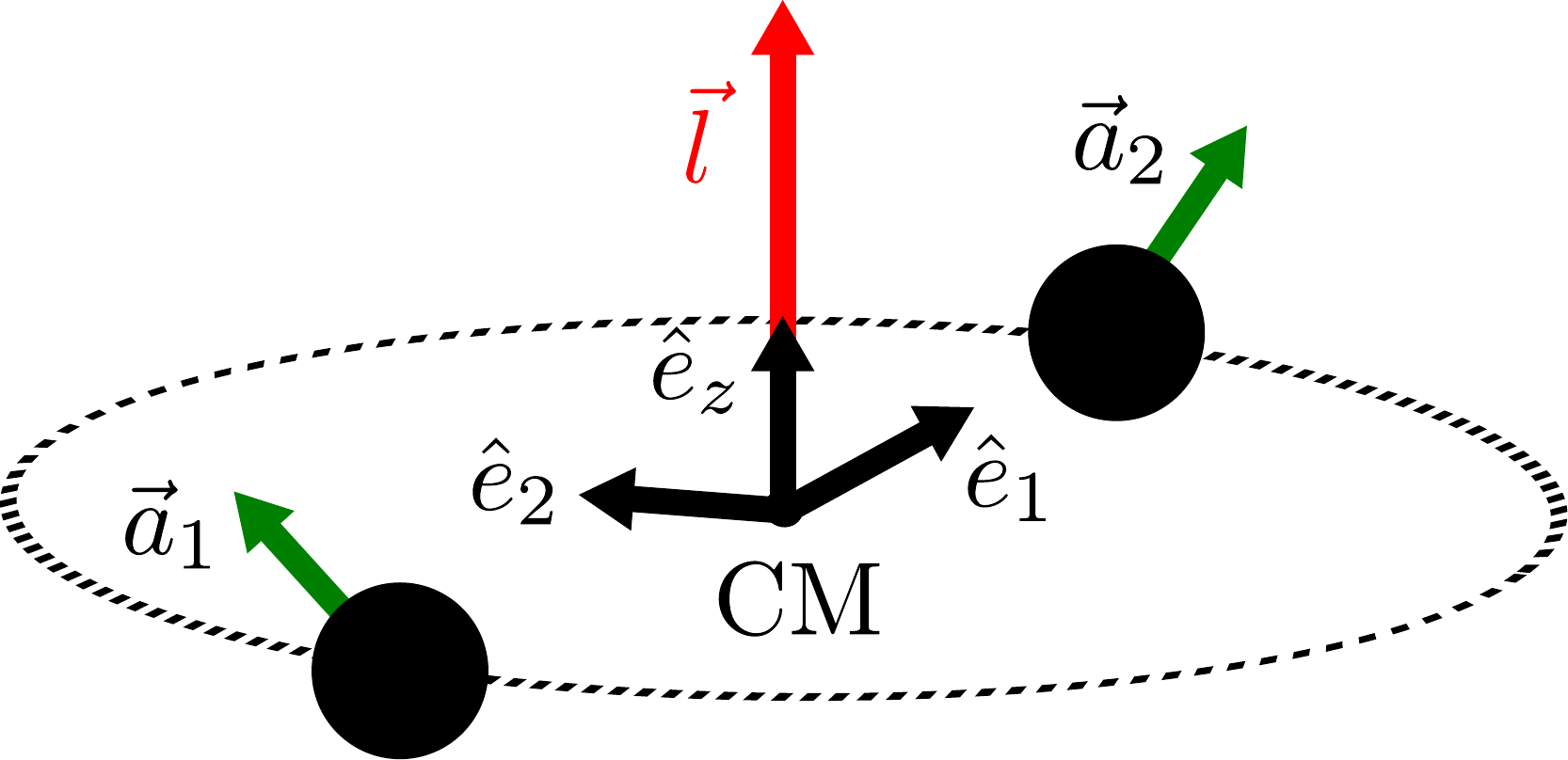}
\caption{\small{Reference system used in the calculation of the spin of the BH remnant and the recoil kick velocity imparted at coalescence time due to asymmetrical emission of gravitational waves.}}

\label{fig:RecoilReferenceSystem}
\end{figure}
%.........................................................................

The second channel of evolution for BH spin corresponds to mergers of BHs. This is a common event in the hierarchical structure formation process, in which two merging galaxies ultimately form a deeper common potential well. After this, the two central BHs approach each through dynamical friction, form a binary, and finally coalescence. During the final inspiral phase, when the orbital angular momentum and energy are driven away by gravitational waves, we track the spin of BH remnants by using the analytical fit of \citet{Rezzolla2008}, which is obtained from relativistic numerical simulations of coalescing binary Kerr BHs. The spin of the BH remnant can be expressed as:
%eeeeeeeeeeeeeeeeeeeeeeeeeeeeeeeeeeeeeeeeeeeeeeeeeeeeeeeeeeeeeeeeeeeeeeeeeeeeee
\eq{FinalSpinBHBH}
{ \bds{a}^f = \frac{1}{(1+q)^2}( \bds{a}_1 + \bds{a_2}q^2 + \bds{\ell}q ), }
%eeeeeeeeeeeeeeeeeeeeeeeeeeeeeeeeeeeeeeeeeeeeeeeeeeeeeeeeeeeeeeeeeeeeeeeeeeeeee
where $\bds{a}_1$ and $\bds{a}_2$ correspond to the spin parameter vectors\footnote{This is a vector pointing in the direction of the BH spin, but with a magnitude given by the dimensionless spin paramenter.} of the more massive and less massive BHs, respectively; $q=M_2/M_1\leq 1$ is the mass ratio of the BH binary; $\bds{\ell} \equiv \bds{\ell}'/(M_1M_2)$, with $\bds{\ell}'$ corresponding to the difference between the orbital angular momentum $\bds{l}$ when both BHs are widely separated from each other; and $\bds{j}_{\sub{rad}}$ is the angular momentum carried away by gravitational waves before coalescence, i.e. $\bds{\ell}' = \bds{l} - \bds{j}_{\sub{rad}}$. 

We use the analytical expression derived by \citet{Rezzolla2008} for the norm of $\bds{\ell}$:
%eeeeeeeeeeeeeeeeeeeeeeeeeeeeeeeeeeeeeeeeeeeeeeeeeeeeeeeeeeeeeeeeeeeeeeeeeeeeee
\begin{eqnarray}
\nonumber
\ell &=& \frac{s_4}{(1+q^2)^2}( a_1^2 + a_2^2q^4 + 2\bds{a}_1 \cdot \bds{a}_2q^2 ) \\
\nonumber
     & & + \left( \frac{s_5\mu + t_0 + 2}{1 + q^2} \right) ( a_1 \cos{\phi_1} + a_2 q^2 \cos\phi_2 ) \\
\label{eqn:Normell}
     & & + 2\sqrt{3} + t_2\mu + t_3 \mu^2,
\end{eqnarray}
%eeeeeeeeeeeeeeeeeeeeeeeeeeeeeeeeeeeeeeeeeeeeeeeeeeeeeeeeeeeeeeeeeeeeeeeeeeeeee
where $\phi_1$ and $\phi_2$ are the angles subtended by the vectors $\bds{a}_1$ and $\bds{a}_2$ with respect to $\bds{\ell}$, respectively. The adopted numerical parameters are: $s_4 = -0.129$, $s_5 = -0.384$, $t_0 = -2.686$, $t_2 = -3.454$ and $t_3 = 2.353$.

In the same fashion as \citet{Rezzolla2008}, we assume the orientation of $\bds{\ell}$ to coincide with that of the total angular momentum of the BH binary before coalescence, i.e. $\bds{l} = \bds{l}_1 + \bds{l}_2$. This is computed directly from the positions and velocities of the two BHs, i.e. $\bds{l}_i = M_{\sub{BH,i}}(\bds{r}_1 - \bds{r}_{\sub{CM}})\times (\bds{v}_i - \bds{v}_{\sub{CM}})$, with $i=1,2$ and CM refering to the centre of mass of the binary. 

Additionally, we also compute the recoil kick velocity imparted to the remnant due to the asymmetrical emission of gravitational waves at coalescence. We use the numerical fitting formula proposed by \citet{Campanelli2007}:
%eeeeeeeeeeeeeeeeeeeeeeeeeeeeeeeeeeeeeeeeeeeeeeeeeeeeeeeeeeeeeeeeeeeeeeeeeeeeee
\begin{eqnarray}
\nonumber
\bds{v}_{\sub{recoil}} &=& v_m \bds{\hat{e}}_1 + v_{\perp}(\cos \xi \bds{\hat{e}}_1 + \sin \xi \bds{\hat{e}}_2 ) + v_{\parallel}\bds{\hat{e}}_z, \\
\nonumber
v_m &=& A \eta^2 \sqrt{ 1 - 4 \eta } (1 + B\eta), \\
\nonumber
v_{\perp} &=& \frac{H \eta^2}{(1+q)}( a_{1}^\parallel - q a_{2}^\parallel ), \\
\label{eqn:vrecoil}
v_{\parallel} &=& K\cos( \Theta - \Theta_0 )\frac{\eta^2}{1 + q}( a_{1}^\perp - q a_{2}^\perp ),
\end{eqnarray}
%eeeeeeeeeeeeeeeeeeeeeeeeeeeeeeeeeeeeeeeeeeeeeeeeeeeeeeeeeeeeeeeeeeeeeeeeeeeeee
where $\bds{\hat{e}}_1$, $\bds{\hat{e}}_2$ are orthogonal unit vectors lying on the orbital plane, and $\bds{\hat{e}}_z$ is perpendicular to it, i.e.~in the direction of the orbital angular momentum (see Figure~\ref{fig:RecoilReferenceSystem}). $\perp$ and $\parallel$ refer to vector components perpendicular and parallel to $\bds{\hat{e}}_z$, $\Theta$ measures the angle between the vector $\bds{a}_1 - q \bds{a}_2$ and $\bds{\hat{e}}_1$, and 
$\eta \equiv q/(1+q)^2$ is the symmetric mass ratio. The adopted numerical parameters in these equations are: $A = 1.2\times 10^4$ km/s, $B = -0.93$, $H = 7.3 \times 10^3$ km/s, $K = 6\times 10^4$ km/s, $\xi = 1.536\ (88^\circ)$ and $\Theta_0 = 0.184$.

%==================================================================================================
\section{Black Hole model}  \label{SecBHmodel}
%==================================================================================================

Modelling BHs in a cosmological context is a challenging task due to many fundamental uncertainties, including the nature of the non-linear coupling between galaxy formation and black hole properties; poorly understood and constrained models of mass accretion onto BHs and AGN-gas interactions; the inherent multi-scale physics of the problem, in which a comparatively small region around the BH impacts significantly largers scales in the host galaxy. These difficulties make an ab-initio
treatment of small-scale BH physics within computational models of galaxy formation impossible with current numerical techniques and resources. Consequently, sub-grid approximations have to be adopted. Specifically, we use the fiducial BH model developed by \citet{Weinberger2017} and references therein. This BH model was adopted in the IllustrisTNG project as well.

%--------------------------------------------------------------------------------------------------
\subsection{Fiducial model}
%--------------------------------------------------------------------------------------------------

Our fiducial default model \citep[based on][]{Weinberger2017} comprises different aspects of BH physics that are crucial for modelling BH-galaxy co-evolution, namely BH growth, seeding of BHs in newly formed proro-galaxies, and feedback associated with BH accretion. In what follows, we describe the modelling of each of these aspects in turn.

%..................................................................................................
\subsubsection{Black hole growth}
%..................................................................................................

BHs gain mass and angular momentum through gas accretion from the ISM and mergers with other BHs. For gas accretion, we adopt the standard Bondi-Hoyle-Lyttleton formula \citep{Hoyle1939,Bondi1944,Bondi1952}:
%eeeeeeeeeeeeeeeeeeeeeeeeeeeeeeeeeeeeeeeeeeeeeeeeeeeeeeeeeeeeeeeeeeeeeeeeeeeeee
\eq{BondiAcc}
{ \dot M_{\sub{Bondi}} = \frac{4\pi G^2 M_{\sub{bh}}^2 \rho}{c_s^3}, }
%eeeeeeeeeeeeeeeeeeeeeeeeeeeeeeeeeeeeeeeeeeeeeeeeeeeeeeeeeeeeeeeeeeeeeeeeeeeeee
where $\rho$ and $c_s$ are the density and sound speed of the gas near the BH, respectively. When the accretion rate of the BH-disc system exceeds the Eddington rate, its value is capped as follows:
%eeeeeeeeeeeeeeeeeeeeeeeeeeeeeeeeeeeeeeeeeeeeeeeeeeeeeeeeeeeeeeeeeeeeeeeeeeeeee
\eq{BHAcrretion}
{ \dot M = \mathrm{min}( \dot M_{\sub{Bondi}}, \dot M_{\sub{Edd}} ), }
%eeeeeeeeeeeeeeeeeeeeeeeeeeeeeeeeeeeeeeeeeeeeeeeeeeeeeeeeeeeeeeeeeeeeeeeeeeeeee
where $\dot M_{\sub{Edd}}\equiv L_{\sub{Edd}}/\epsilon_r c^2$ and $L_{\sub{Edd}}$ is the Eddington luminosity as defined in the spin evolution model. For BH mergers, the masses of the two BHs in the coalescing BH binary are simply added up in the remnant, neglecting mass loss due to gravitational 
wave emission.

%..................................................................................................
\subsubsection{Black hole orbits}
%..................................................................................................

Due to our inability to resolve the local environment of each BH, it is not possible to accurately capture dynamical friction exerted by the background dark matter, gas and stars on the BHs. However, accounting for this process would be important for the correct computation of the BH orbits and the BH binary coalescence timescale. To circumvent this problem, the position of each BH is simply fixed to the local gravitational potential minimum of the host galaxy and its velocity is set to the mean mass-weighted velocity of the region. This has as a consequence that BH binary coalescence cannot be treated in a realistic fashion and a ``numerical'' coalescence must be implemented instead, i.e. two BHs merge once their relative separation is smaller than two softening lengths. Although this approach is simplistic and leads to underestimated BH coalescence timescales and fractions of free floating BHs, it is also numerically robust and prevents BHs to unrealistically migrate out of the centres of their host galaxies.

Note that we have provided an estimate of the recoil velocity in a BH binary coalescence in equation~(\ref{eqn:vrecoil}), however, this has no real impact on the BH population in the present set of simulations due to our special BH positioning scheme, i.e.~BHs cannot recoil by construction. Therefore, we adopt the same approach as \citet{Sijacki2009}, i.e.~we compare the recoil velocity at coalescence time with the escape velocity of the host halo in order to assess the impact of our spin evolution model on the fraction of escaping BHs, which are the result of strong gravitational recoils. Mild gravitational recoils might not be powerful enough to expel a BH from its galaxy, but they can lead to off-centred BHs. As highlighted by \citet{Sijacki2011} and \citet{Blecha2011}, BHs wandering off the galactic centre have a sizeable impact on the BH growth, increasing thus the scatter in the BH mass -- host galaxy relationships. In future work we intend to include a sub-grid model of dynamical friction in order to study the impact of off-centred BHs more realistically \citep[see also][]{Tremmel2015}.

%..................................................................................................
\subsubsection{Black hole seeding}
%..................................................................................................

Although virtually every galaxy hosts a supermassive BH, there is a significant theoretical uncertainty about the origin and early growth of these objects (e.g. \citealt{Volonteri2010}). This makes it difficult to physically motivate models of BH seeding in simulations. Many recent hydrodynamical simulations of galaxy formation employ heuristic approaches, where a halo is seeded with an already massive BH upon fulfilment of a given condition, circumventing thus the phase of early growth. For instance, \citet{Bellovary2011} and \citet{Tremmel2017} seed BHs in locations where the local gas becomes dense, cold and has a low metal content. In the fiducial model of \citet{Weinberger2017} and in other previous works, a BH of mass $M_{\sub{seed}}$ is placed at the position of the potential minimum of any halo that exceeds a total mass $M_{\sub{FOF}}$ and does not contain a BH yet. Despite its simplicity, this method is relatively insensitive to the details of the early BH growth physics. It only assumes that this process is efficient enough to populate all halos with a low-mass supermassive black hole, whether or not this then grows further is a matter of the local conditions around it.

%..................................................................................................
\subsubsection{Black hole feedback}
%..................................................................................................

There is widespread theoretical consensus that there exist at least two types of accretion flows onto BHs (e.g.~\citealt{Begelman2014}, and references therein). One type operates in BHs with high accretion rates and is associated with the formation of classical accretion discs \citep{Shakura1973}. The second type occurs only in BHs with low accretion rates and produces a more spherical and hotter accretion flow \citep{Shapiro1976, Ichimaru1977}. \citet{Weinberger2017} associate these types of accretion with two different BH feedback modes, namely a thermal quasar feedback mode that is activated during the high accretion state, and a kinetic feedback mode that operates in BHs with low-accreting BHs.

In the  quasar mode, the energy liberated by the BH is injected thermally into the surrounding gas and is given by \citep{Springel2005}:
%eeeeeeeeeeeeeeeeeeeeeeeeeeeeeeeeeeeeeeeeeeeeeeeeeeeeeeeeeeeeeeeeeeeeeeeeeeeeee
\eq{QuasarFeedback}
{ \Delta \dot E_{\sub{high}} = \epsilon_{\sub{f, high}} \epsilon_r \dot M c^2, }
%eeeeeeeeeeeeeeeeeeeeeeeeeeeeeeeeeeeeeeeeeeeeeeeeeeeeeeeeeeeeeeeeeeeeeeeeeeeeee
where, as previously defined, $\dot M$ is the accretion rate of the BH-disc system and $\epsilon_r$ the radiative efficiency, which is spin-dependent in our case whereas in the fiducial model a fixed value $\epsilon_r = 0.1-0.2$ is commonly adopted. Finally, the parameter $\epsilon_{\sub{f, high}}$ quantifies the efficiency of the energy coupling process to the local environment.

For the kinetic mode, the liberated energy is 
%eeeeeeeeeeeeeeeeeeeeeeeeeeeeeeeeeeeeeeeeeeeeeeeeeeeeeeeeeeeeeeeeeeeeeeeeeeeeee
\eq{KineticFeedback}
{ \Delta \dot E_{\sub{low}} = \epsilon_{\sub{f, kin}} \dot M c^2, }
%eeeeeeeeeeeeeeeeeeeeeeeeeeeeeeeeeeeeeeeeeeeeeeeeeeeeeeeeeeeeeeeeeeeeeeeeeeeeee
where $\epsilon_{\sub{f, kin}}$ is the coupling efficiency. Unlike the thermal quasar mode, the coupling process is mechanical, i.e.~pure kinetic energy is injected into the gas without changing its temperature. This injection is isotropic when averaged in time, which yields a net conservation of linear momentum. Note that there is no dependence on the BH spin through $\epsilon_r$ as the low accretion state is thought to be radiatively inefficient. A direct spin dependence might, however, be present in reality, depending on the mechanism responsible of delivering the feedback energy in the kinetic mode. For instance, if small-scale jets produced by the Blandford-Znajek mechanism take place, the jet power has a direct dependence on the BH spin \citep{Blandford1977, Yuan2014}:
%eeeeeeeeeeeeeeeeeeeeeeeeeeeeeeeeeeeeeeeeeeeeeeeeeeeeeeeeeeeeeeeeeeeeeeeeeeeeee
\eq{JetPower}
{ L_{\sub{jet}} \propto B^2_{\sub{pol}}M_{\sub{bh}}^2 a^2, }
%eeeeeeeeeeeeeeeeeeeeeeeeeeeeeeeeeeeeeeeeeeeeeeeeeeeeeeeeeeeeeeeeeeeeeeeeeeeeee
where $B_{\sub{pol}}$ is the poloidal magnetic field at the BH horizon. This scenario might also provide a third spin evolution channel as the rotational energy of the BH is drained by the jet. Nevertheless, in the fiducial model the kinetic feedback mode is assumed to be independent on the BH spin. In future work, we intend to return to this point and explore the consequences of including such a dependence, and also of injecting the kinetic feedback in the direction of the BH spin instead of using a random direction.

A remaining issue is the transition between the high and the low accretion modes, or equivalently, between the thermal quasar mode and the kinetic mode. Given the inability to resolve the small-scale accretion onto a BH, such transition cannot be self-consistently accounted for in cosmological simulations. Therefore, more effective approaches must be adopted instead. For instance, \citet{Sijacki2015} and other previous works have used a fixed threshold value of the fraction $\chi\equiv \dot M/\dot M_{Edd}$, i.e.~when $\chi \lesssim \chi_{\sub{th}}$ the low accretion mode is adopted, otherwise the BH is assumed to be in a high accretion state. This idea is extended in the fiducial model, and the threshold value is scaled with BH mass \citep{Weinberger2017}:
%eeeeeeeeeeeeeeeeeeeeeeeeeeeeeeeeeeeeeeeeeeeeeeeeeeeeeeeeeeeeeeeeeeeeeeeeeeeeee
\eq{CanonicalTransition}
{ \chi_{\sub{th}} = \mathrm{min}\left[ \chi_0 \left( \frac{M_{\sub{bh}}}{10^8\ M_{\odot}} \right)^{\beta}, 0.1 \right], }
%eeeeeeeeeeeeeeeeeeeeeeeeeeeeeeeeeeeeeeeeeeeeeeeeeeeeeeeeeeeeeeeeeeeeeeeeeeeeee
with $\chi_0$ and $\beta$ free parameters that are calibrated to reproduce observational trends.

%--------------------------------------------------------------------------------------------------
\subsection{Self gravity in the accretion disc as a switch of BH kinetic feedback}
%--------------------------------------------------------------------------------------------------

The implementation of a kinetic BH feedback mode in the IllustrisTNG simulations has proven to be extremely important to simultaneously account for quenching of star formation and correct gas abundances in high mass galaxies \citep{Weinberger2017, Weinberger2018}. Nevertheless, apart from the notion that the kinetic BH feedback mode occurs in massive galaxies, the exact conditions under which it operates remain unknown. This uncertainty is addressed in the fiducial model by adopting an \textit{ad hoc} feedback switch (equation \ref{eq:CanonicalTransition}) that is calibrated to reproduce observational trends. An important example for the outcome of this is the strong galaxy colour bimodality in IllustrisTNG, which is remarkably consistent with observations \citep{Nelson2018}. This represents a major improvement with respect to the first generation of Illustris simulations, in which the galaxy colour bimodality was much weaker \citep{Nelson2015}. 

In spite of many encouraging successes, there are still some tensions between IllustrisTNG and observations. For example, \citet{Rodriguez-Gomez2018} report a galaxy morphology -- colour relation that is inconsistent with observations, i.e. there is an excess of red discs and blue spheroids in the simulations. This discrepancy might be caused by the limited mass resolution of the simulations, which seems to be insufficient to produce starbursts \citep{Sparre2016}. Given that galaxy mergers trigger starbursts and drive morphological changes, and that post-merger galaxies tend to have a quiescent star forming activity, the inability to produce starbursts might naturally lead to a lacking connection between galaxy morphology and colour. \citet{Rodriguez-Gomez2018} suggest that the missing link is BH gas accretion during galaxy mergers. They argue that the limited mass resolution likely causes an underestimation of merger-induced gas inflows that feed central BHs. This would ultimately lead to an under-powered BH feedback and a potential excess of blue spheroids due to poor quenching. Nevertheless, we note that the BH mass -- stellar mass relation is well reproduced in IllustrisTNG \citep{Weinberger2018}, which means that BHs are not significantly underfed. Therefore, gas accretion is not likely the source of the problem.

An alternative explanation of the tension in the morphology -- colour relation could lie in the use of an inadequate BH feedback switch. Recall that the colour bimodality is well reproduced in IllustrisTNG \citep{Nelson2018}, demonstrating that the kinetic BH feedback mode is an efficient quenching mechanism. However, the inconsistent morphology means that the BH feedback switch is selecting the wrong galaxy population to quench. Thus, spheroids that should be quenched but are not, become blue, and discs that are quenched but should not, become red. The origin of this problem might be related to the \textit{ad-hoc} nature of the fiducial BH feedback switch, and its lack of a physical connection to the galaxy morphology. In this case it would be very difficult, if not impossible, to calibrate equation~(\ref{eq:CanonicalTransition}) to also reproduce the morphology -- colour relation in the simulations.

In order to construct a more physically motivated BH feedback switch, we start from the notion that there are two different BH accretion modes \citep{Begelman2014}. A high accretion mode, in which a radiatively efficient thin disc develops; and a low accretion mode, in which the accretion flow is spherical, hot, and radiatively inefficient. We note that these modes closely resemble the coherent and self-gravity regimes of the accretion disc in the BH spin evolution model (see Section~\ref{SecBHSpins}). The coherent accretion regime is compatible with the formation of a classical Shakura-Sunyaev disc in the high accretion mode, whereas the fragmentation of the disc in the self-gravity regime provides an ideal scenario for the spherical hot accretion flow in the low accretion mode. Following \citet{Weinberger2017}, we associate the quasar thermal feedback mode with the coherent regime (high accretion mode), and the kinetic feedback mode with the self-gravity regime (low accretion mode). Therefore, we can use the self-gravity condition $R_{\mathrm{sg}}\leq R_{\mathrm{warp}}$ as a BH feedback switch. A similar scenario is proposed by \citet{Garofalo2010}, in which merger-induced retrograde accretion onto a BH is much more efficient in producing non-thermal jet outflows. Note that our argument is a mere working hypothesis, meant to explore the still poorly understood physical connection between BH accretion and BH feedback. Nonetheless, our approach does not require calibration, and self-consistently predicts the activation of the kinetic feedback mode based only on the physical properties of BHs, which represents a big improvement with respect to previous \textit{ad-hoc} schemes.

\begin{figure}
\centering
\includegraphics[width=0.5\textwidth]{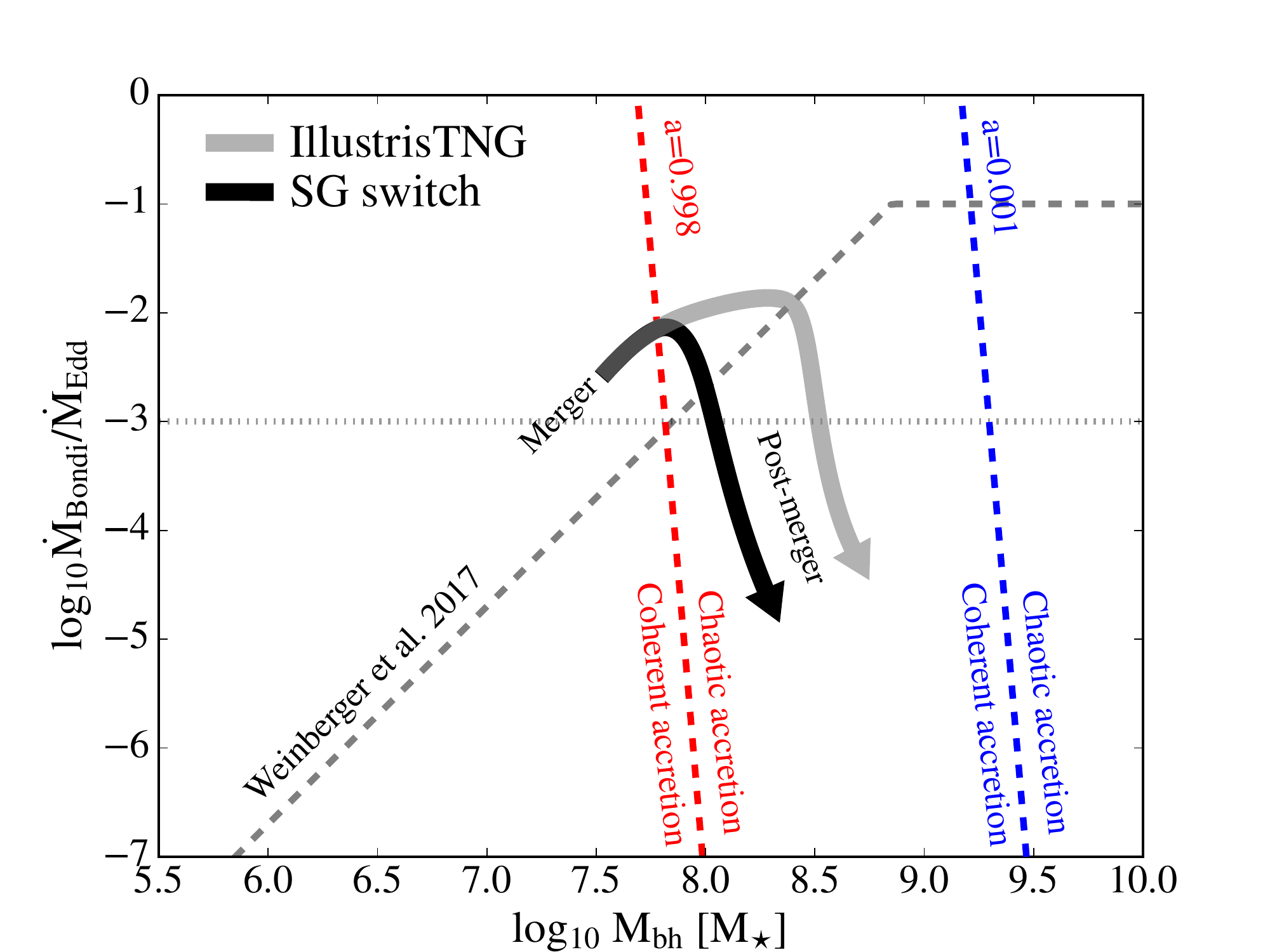}
\caption{ \small Diagram of Eddington ratio vs BH mass illustrating different BH feedback switches. The dashed red ($a=0.998$) and blue ($a=0.001$) lines correspond to the transition between coherent and chaotic accretion based on the self-gravity of the accretion disc, as proposed here. The grey dashed line corresponds to the switch between the quasar and kinetic BH feedback modes adopted by \citet{Weinberger2017}.}
\label{fig:DiagramEvol}
\end{figure}

Starting from the condition $R_{\mathrm{sg}}\leq R_{\mathrm{warp}}$, we can derive a parametric equation for the mass threshold $M_{\mathrm{bh, th}}(\dot{M}, a)$ at which the transition from the coherent regime to the self-gravity regime occurs. In order to do so, we use equations (\ref{eq:Rwarp}) and (\ref{eq:Rselfgrav}) for the warp radius $R_{\mathrm{warp}}(M_{\mathrm{bh}},\dot{M}, a)$ and the self-gravity radius $R_{\mathrm{sg}}(M_{\mathrm{bh}},\dot{M}, a)$, respectively. In Figure~\ref{fig:DiagramEvol}, we show the BH mass -- Eddington ratio diagram with two different mass thresholds for two extreme values of the spin parameter, i.e. $M_{\mathrm{bh, th}}(\dot{M}, a=0.998)$ and $M_{\mathrm{bh, th}}(\dot{M}, a=0.001)$. For comparison purposes, we also show the mass threshold associated to the BH feedback switch of the fiducial model \citep[we use $\chi_0=0.002$ and $\beta=2.0$ as in][]{Weinberger2017}.

In this figure, we also sketch the evolutionary path of a maximum spinning BH of mass $\sim 10^8\ M_{\odot}$\footnote{According to the BH mass -- stellar mass relation \citep{Kormendy2013}, a BH mass of $\sim 10^8\ M_{\odot}$ corresponds to a galaxy stellar mass of $\sim 10^{10}\, {\rm M}_{\odot}$, which is roughly the mass scale above which quenching becomes efficient and galaxies migrate to the red sequence.} during a galaxy merger. The grey arrow represents what happens if the fiducial BH feedback switch is adopted. Likewise, the black arrow  represents the BH evolution if the self-gravity switch is used instead. Considering that, during a galaxy merger, the central BH experiences a boosted growth due to gas accretion and/or coalescence with a second BH, both quantities, BH mass and Eddington ratio, will unavoidably increase. This sets a specific direction of evolution in the diagram. Adopting the self-gravity switch will clearly favour quenching of post-merger galaxies compared to the fiducial switch. This, in conjunction with merger-induced morphological changes, might help recovering the colour-morphology relation in the simulation. We also note that galaxies hosting BHs with lower spin values will be quenched at higher stellar masses.

%.........................................................................
%	Figure 4
%.........................................................................
%.........................................................................

Some high redshift quasars exhibit extremely large BH masses ($M_{\mathrm{bh}}\gtrsim 10^9\, {\rm M}_{\odot}$) and high accretion rates. The kinetic feedback mode should clearly not be operating under these conditions as it would expel most of gas content of the host galaxy, so the thermal quasar mode must be operating instead. The fiducial model solves this by specifying an upper limit for the Eddington rate ($\dot{M}/\dot{M}_{\mathrm{Edd}}=0.1$) for the kinetic feedback mode to operate.

For the self-gravity switch, a self-consistent solution might instead emerge out of the spin evolution model. Once the self-gravity regime is reached, the kinetic feedback mode is activated and most of the central gas is heated and/or pushed away. This results in a significant decrease of the BH accretion rate and a suppression of the gas evolution channel. After this, only BH binary coalescence can alter the spin. Nevertheless, if a gas-rich event manages to substantially increase the BH accretion rate again, the BH will quickly be spun down, thereby increasing the BH mass threshold of the self-gravity switch and activating again the thermal quasar mode. Additionally, if a spin dependent kinetic feedback based on the Blandford-Znajek mechanism were operating (equation~\ref{eq:JetPower}), a BH with a decreasing spin would weaken the feedback energy as $L_{\sub{jet}}\propto a^2$, thereby aiding further accretion. The feasibility of this scenario is yet to be tested in full; the low abundance of these objects demands simulations of large cosmological volumes at at least moderate resolutions, which is out of the scope of this work.
  
%==================================================================================================
\section{Numerical Set-up}  \label{SecNumerics}
%==================================================================================================

%--------------------------------------------------------------------------------------------------
\subsection{Simulation Code}
%--------------------------------------------------------------------------------------------------

We run cosmological simulations to self-consistently track the spin evolution of supermassive black holes along with other quantities of interest such as BH mass accretion rate, BH feedback and properties of the host galaxies.  We use the $N$-body magneto-hydrodynamics moving-mesh code {\small AREPO} \citep{Springel2010} to solve the non-linear system of differential equations describing gas hydrodynamics and gravitational interactions of dark matter and baryons in such simulations. One of the unique features of {\small AREPO} is that the mesh continuously transforms and flows along with the fluid, unlike in standard grid-based methods. This is achieved through a dynamic unstructured mesh that is constructed as the Voronoi tessellation of a set of mesh-generating points that move with the local fluid velocity. This ensures that every cell approximately maintains a given target mass, and in this fashion, regions of high density are resolved automatically with smaller cells than low density regions, which is crucial for modelling processes such as gas accretion onto BHs residing in high density galaxy centres. Furthermore, because the mesh moves with the fluid, the advection errors inherent in Eulerian mesh codes are strongly reduced. In fact, a discretization that is manifestly Galilean invariant is achieved in this way.  Also, there are no preferred coordinate directions in this scheme, avoiding potential spurious alignments of gas structures with the grid, as it can happen in Cartesian mesh codes when the resolution is low. This helps in accurately tracking the co-evolution of the spin of BHs and the angular momentum of the gas surrounding BHs in an isotropic way.

Gravitational forces in {\small AREPO} are calculated using the Tree-PM scheme \citep{Xu95}, in which long-range forces are computed with a particle-mesh method and short-range forces are followed via a hierarchichal octree algorithm \citep{Barnes86}. Apart from gravitational interactions and hydrodynamics, the code incorporates a complete physics model that includes: gas cooling of primordial and metal lines; a sub-resolution ISM model based on the two phase medium \citep{Springel2003}, in which stars are formed in a stochastic fashion above a density threshold of $0.13\,$cm$^{-3}$; stellar evolution; chemical enrichment; gas recycling; kinetic stellar feedback driven by SNe and black hole physics (section~\ref{SecBHmodel}).

%--------------------------------------------------------------------------------------------------
\subsection{Numerical implementation of spin evolution}
%--------------------------------------------------------------------------------------------------

As discussed in Section~\ref{SecBHSpins}, the BH spin parameter $a$ can be computed in an iterative fashion through equation~(\ref{eq:SpinEvol}); however, an initial condition $a_0$ is still required. Hence, we assign an initial spin value $a_0$ to every BH seed. Once initialised, if a BH seed accretes gas above $1\%$ of its Eddington rate, the gas evolution channel is activated. In what follows, we discuss how to integrate BH spin evolution through gas accretion into the {\small AREPO} code.

The density and the direction of the angular momentum of the gas around BHs are needed to compute the mass accretion rate (equation \ref{eq:BondiAcc}) and the direction of the BH spin (equation \ref{eq:TotalJ}). In order to estimate these properties, we use a spherical SPH kernel $w(r;h)$ around each BH that encloses a prescribed number of gas cells $n_{\mathrm{ngb}}$ in a radius $h$:
%eeeeeeeeeeeeeeeeeeeeeeeeeeeeeeeeeeeeeeeeeeeeeeeeeeeeeeeeeeeeeeeeeeeeeeeeeeeeee
\eq{KernelNumber}
{ n_{\mathrm{ngb}}\approx \sum_i \frac{4\, \pi\, h^3\, m_i}{3\, m_{\mathrm{baryon}}}w(r_i). }
%eeeeeeeeeeeeeeeeeeeeeeeeeeeeeeeeeeeeeeeeeeeeeeeeeeeeeeeeeeeeeeeeeeeeeeeeeeeeee
Here, $m_{\mathrm{baryon}}$ is the target mass of a gas cell and $i$ runs over the cells within the kernel radius.

One important issue that needs to be addressed concerns the time integration of the spin evolution. On the one hand, gravitational interactions and gas hydrodynamics fix the evolution time step $\Delta t$ for every particle/gas cell of the simulation \citep{Springel2010}. For BHs, this time step determines how often properties such as $M_{\mathrm{bh}}$, $\dot{M}$ and the direction of the angular momentum of neighbouring gas are updated. On the other hand, the BH spin is updated after an accretion timescale $t_{\mathrm{acc}}$ has elapsed in the coherent regime, or after the disc has accreted a mass $M_{\mathrm{sg}}$ in the self-gravity regime. This implies that processes such as BH growth and BH feedback, which are governed by $\Delta t$ in the code, are not necessarily synchronised with the evolution of BH spin. This leads to four different cases:

\begin{itemize}
 \item[\textit{(i)}] \textit{BH in coherent regime, $t_{\mathrm{acc}}<\Delta t$:} in this case, several accretion episodes occur within the same time step. At the beginning of the time step, we store the BH properties needed for the spin model, and they are used for every accretion episode occurring within that time step. We also set a cumulative time counter that keeps track of how many accretion episodes have been completed. We stop once the counter equals $\Delta t$ and proceed to the next time step.
 
 \item[\textit{(ii)}] \textit{BH in coherent regime, $t_{\mathrm{acc}}>\Delta t$:} in this case, several times steps elapse in the same accretion episode. We set a time counter at the beginning of the accretion episode that keeps track of how many time steps have been completed. We stop once the counter equals $t_{\mathrm{acc}}$ and proceed to the next accretion episode. The input BH properties for the spin model are averaged over the elapsed time steps.
 
 \item[\textit{(iii)}] \textit{BH in self-gravity regime, $M_{\mathrm{sg}}<\dot M \Delta t$:} this case is similar to \textit{case (i)}, i.e. several accretion episodes occur within the same time step. Given that the BH spin must be updated once a mass $M_{\mathrm{sg}}$ is swallowed by the BH, we set, at the beginning of the time step, a cumulative mass counter that keeps track of how many times this has happened. We stop once the mass counter equals $\dot M \Delta t$ and proceed to the next time step. The BH properties are assumed constant for all the accretion episodes in this time step.
 
 \item[\textit{(iv)}] \textit{BH in self-gravity regime, $M_{\mathrm{sg}}>\dot M \Delta t$:} this case is similar to \textit{case (ii)}. At the beginning of the accretion episode, we set a cumulative mass counter that keeps track of how much mass is accreted in every time step. We stop once the mass counter equals $M_{\mathrm{sg}}$ and proceed to the next accretion episode. The BH properties are averaged over the elapsed time steps.
 
\end{itemize}

At the beginning of every accretion episode, we compute $R_{\mathrm{warp}}$, $R_{\mathrm{sg}}$, $t_{\mathrm{acc}}$ and $M_{\mathrm{sg}}$. This allows us to assess which of the previous cases applies. At the end of an accretion episode, we update the BH spin.

The second evolution channel, BH binary coalescence, can be easily integrated into the code as BH mergers are treated in an instantaneous fashion. This makes it straightforward to apply equation~(\ref{eq:FinalSpinBHBH}) to compute the spin of the BH remnants.

%--------------------------------------------------------------------------------------------------
\subsection{Simulation set}
%--------------------------------------------------------------------------------------------------

We run a set of 5 cosmological simulations in a periodic box with side length of $25\, h^{-1}\ \mathrm{Mpc}$. This is a relatively small size compared to state-of-art cosmological simulations like Illustris \citep{Vogelsberger2014c}, IllustrisTNG \citep{Pillepich2018,Springel2018} or Eagle \citep{Schaye2015}; however, we still form around $1200$ BHs, with roughly $60$ of them being more massive than $10^{8}\, \mathrm{M}_{\odot}$. This makes our simulations suitable to sample the evolution of BH spin in the coherent and self-gravity regimes. For simulation 1, we do not use our BH spin evolution model as this is the control simulation used for comparison purposes. For simulations 2, 3 and 4, we use our self-consistent BH spin evolution model with different values of the parameter $k$. BH feedback and BH growth are affected by BH spin owing to the spin-dependent radiative efficiency of the accretion disc. Besides, we use the fiducial feedback switch as we want to isolate the effects of BH spin evolution. We show the results of these three simulations in Section~\ref{SecResultsStatistics}. Finally, simulation 5 is run with the BH spin model as well, but additionally, we use our BH feedback switch based on the self-gravity of the accretion disc. The results of this simulation are discussed in Section~\ref{SecBHSpinFeedback}. We summarise the model choices for our simulations in Table~\ref{tab:simulations}.

%.........................................................................
\begin{table}
\centering
\begin{tabular}{c | c | c | c}
\textbf{Simulation}&  \textbf{Spin model} & $k$ & \textbf{Feedback switch}\\ \hline
1   &  No & - & Fiducial \\
2   & Yes & 0 & Fiducial \\
3   & Yes & 2 & Fiducial \\
4   & Yes & 10 & Fiducial \\
5   & Yes & 0 & Self-gravity \\ \hline
\end{tabular}
\caption{Overview of our primary cosmologial simulation models. For all the simulations, we employ the physics model of IllustrisTNG with some modifications with respect to our new spin model and treatment of the BH feedback switch. Simulation 1 is used as a control for comparison purposes and corresponds to the unmodified IllustrisTNG model.}
\label{tab:simulations}
\end{table}
%.........................................................................

For all our simulations, we adopt a $\Lambda$CDM cosmology consistent with the Planck intermediate results \citep{PlanckCol2016A}, given by $\Omega_M = 0.3089$, $\Omega_{\Lambda} = 0.6911$, $\Omega_b = 0.0486$, $h = 0.6774$ and $\sigma_8 = 0.8159$. Initial conditions are generated by using an \citet{Eisenstein1998} power spectrum starting at redshift $z = 127$. The initial conditions contain $512^3$ dark matter particles and the same number of gas cells. This corresponds to a mass resolution of $8.4\times 10^6h^{-1}\,\mathrm{M}_{\odot}$ and $1.5\times 10^6h^{-1}\,\mathrm{M}_{\odot}$ for dark matter particles and gas cells, respectively, which is similar to the highest resolution of Illustris TNG100 \citep{Pillepich2018}. The softening length for gas cells, dark matter and star particles is $1$ comoving kpc with a maximum value of $0.5$ proper kpc. For the black hole model, we adopt the following parameters: $M_{\mathrm{FOF}} = 5\times 10^{10}h^{-1}\,\mathrm{M}_{\odot}$, $M_{\mathrm{seed}} = 8\times 10^5 h^{-1}\,\mathrm{M}_{\odot}$,
$\epsilon_{\mathrm{f},\, \mathrm{high}} = 0.1$, $\epsilon_{\mathrm{f},\, \mathrm{kin}} = 0.2$, $\epsilon_r=0.2$ (only for fiducial run), $\chi_0 = 0.002$, $\beta = 2.0$ and $n_{\mathrm{ngb}} = 128$. For the spin evolution model, we use $\alpha = 0.1$ and $a_0 = 0$.

%==================================================================================================
\section{Black hole spin in cosmological simulations}  \label{SecResultsStatistics}
%==================================================================================================

%--------------------------------------------------------------------------------------------------
\subsection{Individual histories}
%--------------------------------------------------------------------------------------------------

\begin{figure*}
\centering
\includegraphics[trim={0cm 0cm 0cm 0cm},width=1.0\textwidth]
{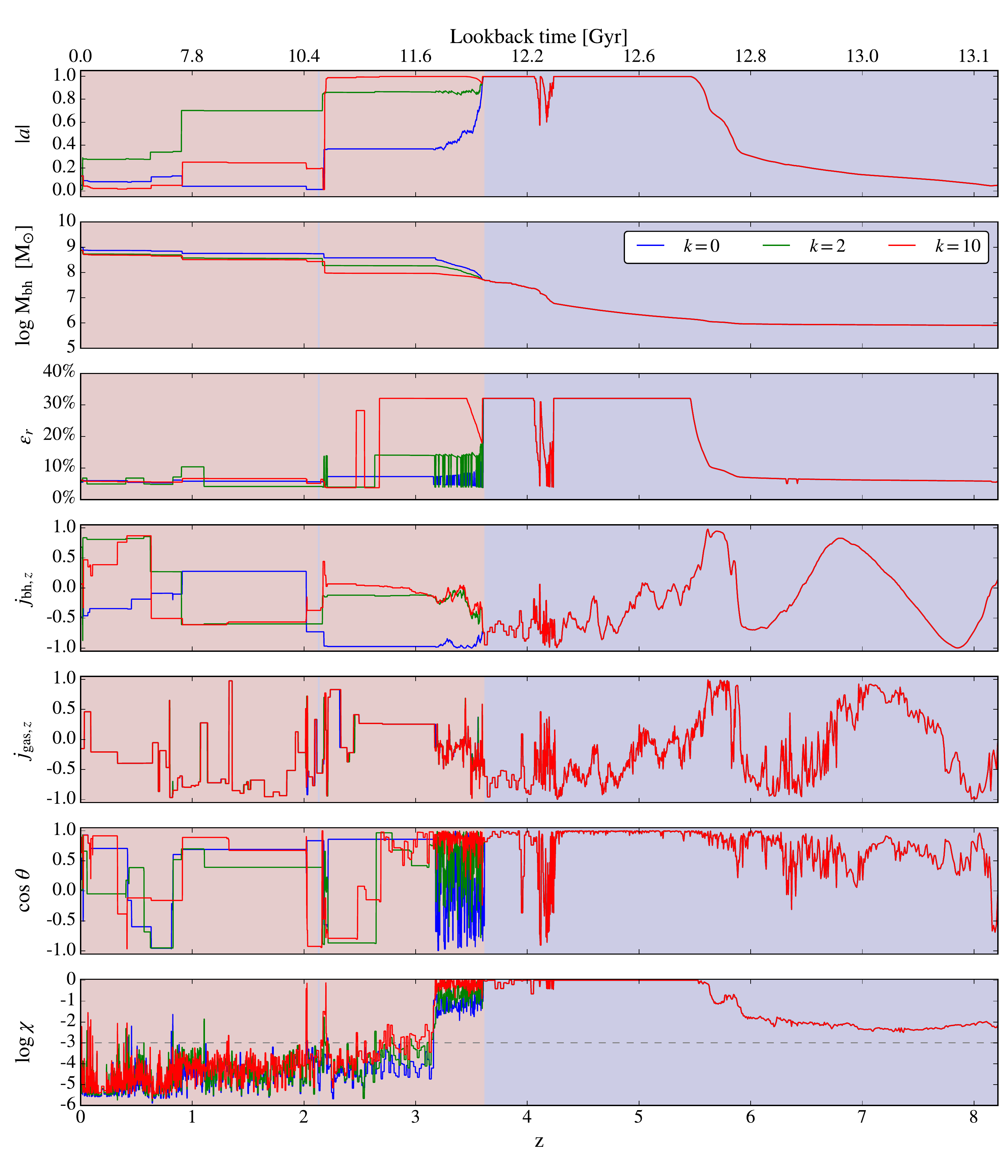}
\caption{\small{Example of the evolution of several BH properties for one of the BHs in our simulations. We track the evolution of this BH for different choices of the concentration parameter $k$, which quantifies the degree of anisotropy in the feeding process. The BH masses at $z=0$ in the different models lie in the range $10^{8.8} - 10^9\,{\rm M}_{\odot}$. \textit{From top to bottom:} BH spin evolution, BH mass evolution, radiative efficiency, Cartesian $z$-component of the normalised BH spin vector, Cartesian $z$-component of the normalised angular momentum vector of the surrounding gas, cosine of the angle between these two vectors, and finally, the Eddington ratio. The $z$-axis corresponds to the fixed reference frame of the simulation. The \textit{blue shade} and the \textit{red shade} represent the coherent and self-gravity accretion regimes, respectively.}}

\label{fig:HistoryBH1}
\end{figure*}

\begin{figure*}
\centering
\includegraphics[trim={0cm 0cm 0cm 0cm},width=1.0\textwidth]
{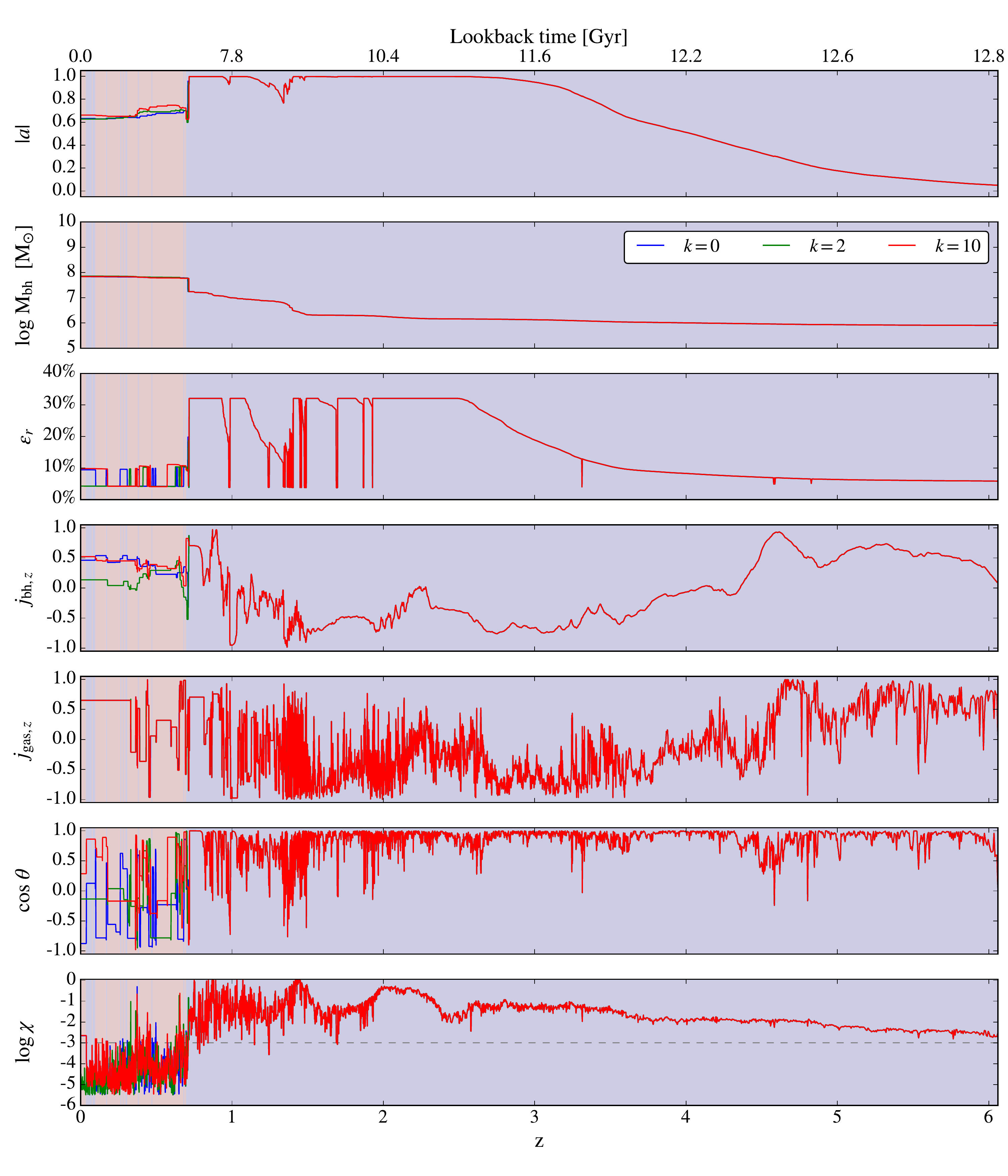}
\caption{\small{Illustrative evolution example of a BH of mass $10^{7.8}\,{\rm M}_{\odot}$ with a more quiescent history. The properties shown are the same as in Fig.~\ref{fig:HistoryBH1}. }}
\label{fig:HistoryBH2}
\end{figure*}

Before discussing the main demographic results of the BH population in our simulations,
it is instructive to examine individual BH histories and how they relate to the spin evolution. In Figures~\ref{fig:HistoryBH1} and \ref{fig:HistoryBH2}, we show two representative cases of the BH evolution. The first example is a very massive BH of mass $10^{8.8}-10^9\,{\rm M}_{\odot}$\footnote{We report a range of BH masses because the anisotropy degree $k$ of the feeding process is varied in each simulation, which yields slightly different mass values even though we show the same black hole in each case.} residing in a galaxy of stellar mass $1.1\times 10^{11}\,{\rm M}_{\odot}$. This BH is seeded at a redshift $z = 8.3$ (lookback time of $13.12\ \mathrm{Gyr}$) when its host halo reaches the threshold mass $M_{\mathrm{FOF}}$. The BH survives until $z=0$, i.e.~it does not merge with any more massive BH throughout its history. 

The evolution of the spin parameter is shown in the top panel of Fig.~\ref{fig:HistoryBH1}. Once the BH is seeded, we find that gas accretion takes about $500\ \mathrm{Myr}$ to spin up the BH to the maximum spin value. Especially in the last $100\ \mathrm{Myr}$ of this period, from lookback time $12.8\ \mathrm{Gyr}$ to $12.7\ \mathrm{Gyr}$, the spin parameter increases very quickly, which is due to the accretion disc being in the coherent regime and the BH reaching the Eddington accretion rate (bottom panel). This is also accompanied by a rapid increment of the radiative efficiency from $\sim 5\%$ to $\sim 30\%$ (third panel), which represents an enhancement of a factor of six in the supplied feedback energy per unit of accreted mass.

%.........................................................................
%	Figure 5
%.........................................................................
%BH1
%.........................................................................

For this BH, the coherent accretion regime extends up to $z=3.6$ (blue shade). During this time, every accretion episode is by construction aligned with the surrounding gas of the BH. Although this type of accretion is more stable than the chaotic counterpart, the orientation of the gas angular momentum can still fluctuate considerably, which is seen in the evolution of the Cartesian $z$-component of the angular momentum $j_{\mathrm{gas},z}$ (fifth panel). Owing to the high mass accretion rate of the disc in the coherent regime, the angular momentum transferred to the BH in every accretion episode is sufficient to keep the BH spin mostly oriented with the gas. This explains why the $z$-projection of the BH spin, $j_{\mathrm{bh},z}$, mimics the global behaviour of $j_{\mathrm{gas},z}$ (fourth panel), and why the angle $\theta$ is small (sixth panel). Small fluctuations in $\bds j_{\mathrm{gas}}$ cannot be followed up by the BH spin as they occur on timescales smaller than the accretion timescale $t_{\mathrm{acc}}$, which in turn has a lower limit imposed by the Eddington rate at a given BH mass and spin, i.e.~corresponding to $\lambda = 1$ in equation~(\ref{eq:Tacc}). 

Between $z=4.3$ and $z=4$, the host galaxy experiences a merger with a smaller galaxy (stellar mass ratio 1:4). This event disturbs the central gas considerably and is reflected in the rapid and violent fluctuations in $j_{\mathrm{gas},z}$ \citep{Mayer2007}. Although the BH is already accreting at its Eddington rate during this time, in this particular case the BH spin is not reoriented fast enough to keep up with the gas fluctuations. This creates an interesting situation in which the BH spin is suddenly misaligned, and counter-rotating accretion takes place even in the coherent regime, i.e.~for $\cos\theta<0$. The spin parameter drops to lower values as the BH is spun down, and thus the radiative efficiency decreases to $5\%$ as well. Due to the self-consistent nature of our model, the BH mass is also impacted during this time and its growth is temporarily boosted (second panel). Note that this behaviour cannot be accounted for in a static model with constant $\epsilon_r$. We remark that these fluctuations of the gas angular momentum are resolved in the simulation and driven by local turbulence \citep{Dubois2014b}. Therefore, they are different than those produced by internal processes in the accretion disc during the self-gravity regime. Once the merger has completed, the central gas becomes stable again and the BH spin rapidly reaches the maximum value.

In general, during the coherent regime the orientation of the BH spin tends to be more stable over time than the angular momentum of the central gas, which allows to interpret the former as a time average of the latter. This might have interesting consequences for the alignment of the BH spin with the host galaxy and the cosmological environment as several processes funnel gas towards the galactic centre from galactic and extra-galactic reservoirs. We defer an analysis of this point to future work.

The self-gravity regime kicks in at $z=4.6$ once the condition $R_{\mathrm{sg}} \leq R_{\mathrm{warp}}$ is satisfied. At this point, the concentration parameter $k$ starts to considerably impact the evolution of the BH. For a fully chaotic accretion ($k=0$), the BH spin does not follow the gas angular momentum anymore, and thus its magnitude rapidly decreases as counter-rotating accretion episodes become very frequent. This reduces the radiative efficiency of the disc and enhances mass growth. For a very coherent accretion ($k=10$), the BH behaves similarly as during the coherent regime. Medium concentrations (e.g.~$k=2$) yield a somewhat intermediate behaviour, i.e.~the BH is fairly aligned with the gas angular momentum, but counter-rotating accretion episodes can still occur occasionally. At $z=3.2$, the kinetic feedback mode is activated and the central gas is quickly heated up and swept away, hence the gas accretion channel is suppressed and only BH binary mergers can significantly alter the BH mass and spin. We note that during the self-gravity regime, $j_{\mathrm{gas},z}$ appears to display a discontinuous behaviour. This is, however, a spurious effect due our recording of all relevant BH properties only when the BH spin is modified, which is done for computational reasons.

%.........................................................................
%	Figure 6
%.........................................................................
%BH2
%.........................................................................

Finally, we show in Figure~\ref{fig:HistoryBH2} a second example of a more quiescent BH of mass $10^{7.8}\,{\rm M}_{\odot}$ formed at $z=6.05$, residing in a galaxy of stellar mass $1.2\times 10^{10}\,{\rm M}_{\odot}$. The evolution of this BH is in part very similar to the previous one, e.g., during the coherent regime, the spin parameter eventually reaches the maximum allowed value while the BH spin is, with a few exceptions, mostly oriented with the gas angular momentum. Nevertheless, as this BH almost never reaches its Eddington rate, its whole evolution is considerably slower. A very interesting aspect is that a major galaxy merger at $z=0.7$ (stellar mass ratio 1:2) triggers both the self-gravity regime and the kinetic feedback mode at the same time, hinting thus a possible connection between these two processes. We explore this hypothesis more thoroughly in Section~\ref{SecBHSpinFeedback}.

%--------------------------------------------------------------------------------------------------
\subsection{Black Hole demographics}
%--------------------------------------------------------------------------------------------------

\subsubsection{Black hole spin distribution}

\begin{figure}
\centering
\includegraphics[trim={0cm 0cm 0cm 0cm},width=0.52\textwidth]
{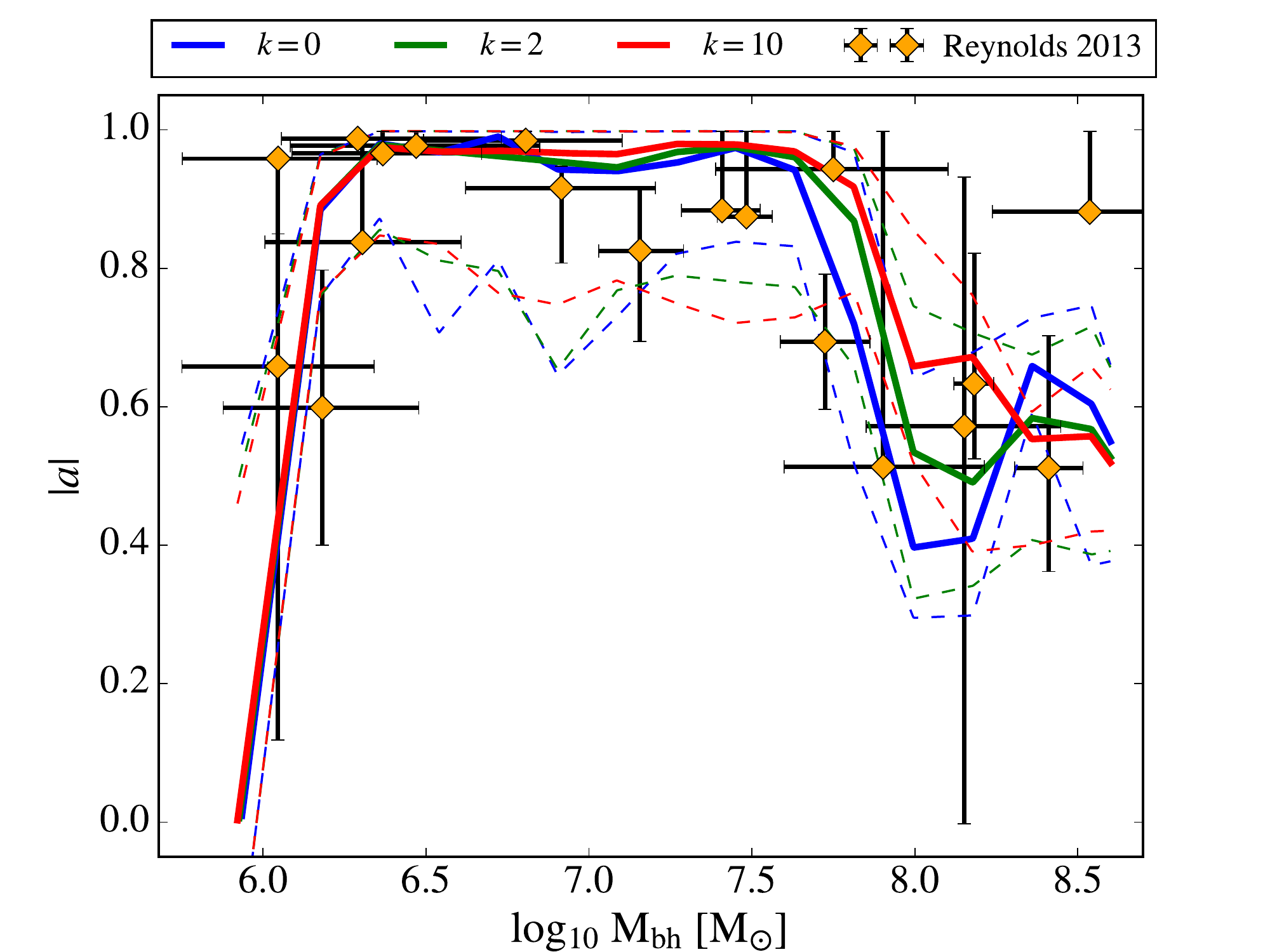}
\caption{\small{Median of the absolute BH spin value distribution as a function of BH mass at $z=0$ (\textit{solid lines}) for three different values of the concentration parameter $k$ that quantifies the anisotropy of the feeding process. \textit{Dashed lines} represent $25\%$ and $75\%$ percentiles. \textit{Orange diamonds} with $1\sigma$ bars are observational data compiled by \citet{Reynolds2013}.}}

\label{fig:SpinDistribution}
\end{figure}

Having shown how the spin model works in individual BHs, we proceed to study the demographics of the BH population in our simulations. We start by analysing the distribution of the absolute value of the BH spin as a function of BH mass. We plot in Figure~\ref{fig:SpinDistribution} the results for simulations 2, 3 and 4, which include the spin model, but which use different values for the concentration parameter $k$. 

One common result is the initial monotonic growth of the spin for BHs with masses of $10^6 - 10^{6.5}\,{\rm M}_{\odot}$, i.e.~right after they are seeded. The small dispersion during this first phase indicates that all BHs follow the same path, i.e., at the beginning, BH spin evolution depends mainly on BH mass, no matter how fast the mass is accreted. This behaviour is explained by the explicit mass dependence in equation~(\ref{eq:SpinEvol}), and the ease with which the spin of a low-mass BH can be reoriented during an accretion episode, i.e.~counter-rotating accretion is very unlikely to occur and the BH always gets spun up. For masses larger than $10^{7.5}\,{\rm M}_{\odot}$, the spin decreases to intermediate values and the distribution becomes more homogeneous. This is caused by BH binary coalescence and BHs entering the self-gravity regime, in which the frequency of counter-rotating accretion episodes is controlled by the concentration parameter. Approximately at this mass scale, the kinetic feedback mode kicks in as well, which limits gas accretion and leaves BH coalescence as the only effective spin evolution channel. This also explains why the different choices of the concentration parameter do not affect the spin distribution considerably, i.e.~the gas evolution channel is briefly active during the self-gravity regime. From now on, we adopt for definiteness in all our analysis the chaotic accretion in the self-gravity regime as the default mode, i.e.~we use only the simulations with $k=0$.

Although direct observational measurements of BH spins are difficult and challenging to obtain, there have been several attempts made through X-ray spectroscopy of the iron fluorescence line of the accretion disc. \citet{Reynolds2013} made a compilation and a quality assessment of the measurements, which is necessary due to several contradictory results. We include these data in Figure~\ref{fig:SpinDistribution} (orange diamonds) and compare them with our BH spin distribution. We find a very good agreement, i.e.~spins tend to have a high value at intermediate BH masses, but they decrease for more massive BHs. Our results are also consistent with previous numerical studies of spin evolution in a cosmological context \citep[e.g.][]{Dotti2013, Dubois2014c}.

%.........................................................................
%	Figure 7
%.........................................................................
%Spin distribution at z=0
%.........................................................................

In order to test the robustness of our results, we have run simulations with different initial spin values, namely $a_0 = 0.5$ and $a_0 = 0.9$. In all cases, the spin distributions for masses larger than $10^{6.5}\,{\rm M}_{\odot}$ were almost identical, thereby demonstrating the insensitivity of the model on the initial conditions. This can be easily explained by the fact that BH binary mergers become common for massive BHs, and thus they rapidly erase any trace of the initial conditions and introduce a large dispersion in the spin distribution. Counter-rotating accretion also aids this as it becomes increasingly more likely to occur and is a relatively stochastic process.

\subsubsection{Black hole mass -- stellar mass relation}

\begin{figure*}
\centering
\includegraphics[width=1\textwidth]{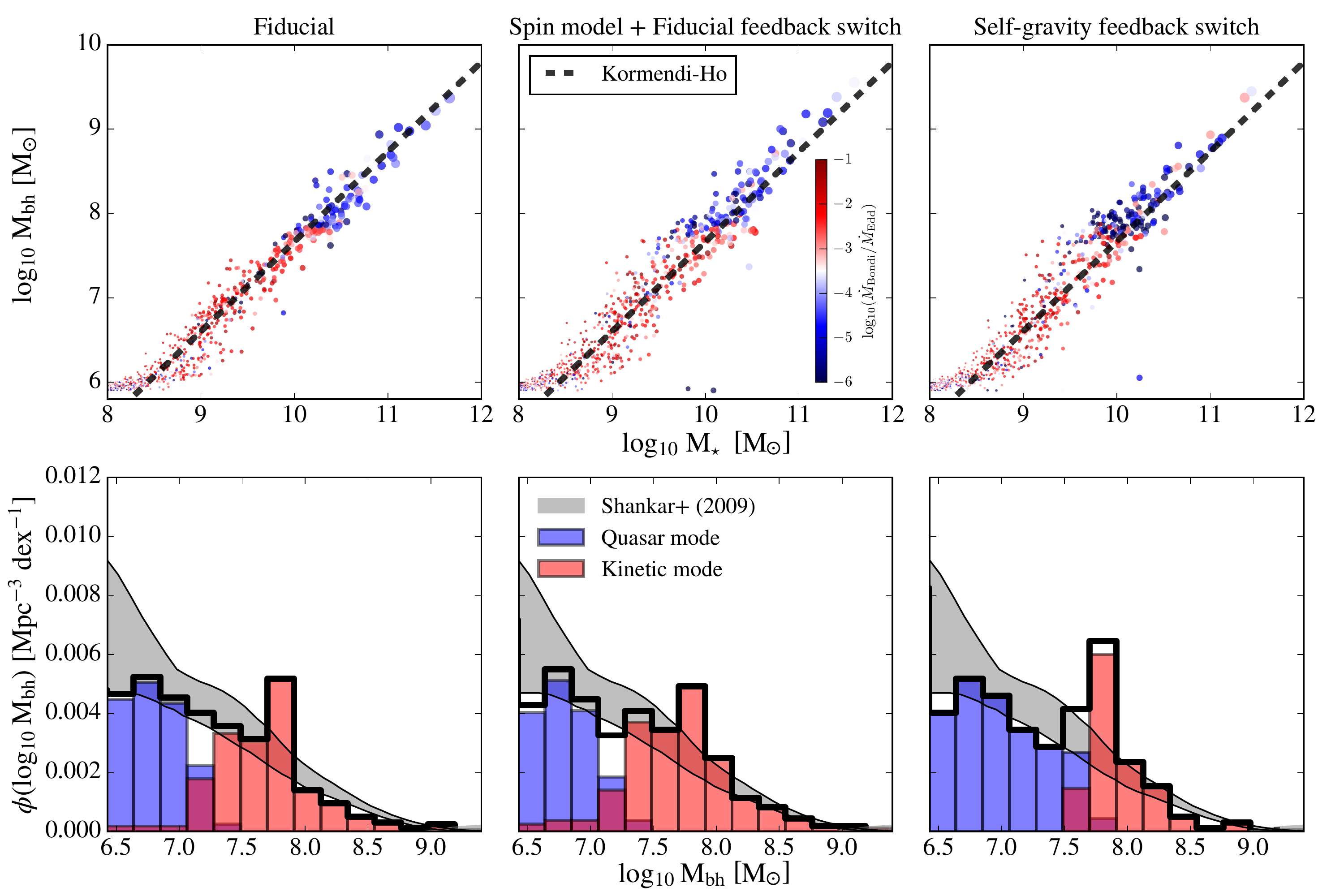}
\caption{ \small BH mass -- stellar mass relation (\textit{top panels}) and BH mass function (\textit{bottom panels}) at $z=0$ for three different models, namely the fiducial model (\textit{left panels}), our spin model with the fiducial feedback switch (\textit{central panels}) and our full model that includes the self-gravity feedback switch (\textit{right panels}).
For the \textit{top panels}, colour encodes the Eddington rate and the size of the symbols is scaled with stellar mass to increase visibility. Stellar masses are computed from all star particles within twice the stellar half mass radius ($2\, R_{1/2,\star}$). The \textit{black dashed line} corresponds to an observational fit taken from \citet{Kormendy2013}. For the \textit{bottom panels}, the blue histogram represents BHs that have been always in the quasar feedback mode while the red histogram represents BHs that are or have been in the kinetic feedback mode at some point during their history, even if they are currently in the quasar feedback mode. The \textit{grey shade} corresponds to observational constrains derived by \citet{Shankar2009}.}
\label{fig:BHMassRelation}
\end{figure*}

Amongst the achievements of the IllustrisTNG project and its kinetic BH feedback mode is the reproduction of the BH mass -- stellar mass relation \citep{Weinberger2017, Weinberger2018}. Given that our spin model has a direct impact on BH accretion and the associated feedback through the spin-dependent radiative efficiency, we investigate how it affects the BH mass -- stellar mass relation and the BH mass function. These are shown in Figure~\ref{fig:BHMassRelation}, top and bottom panels, respectively. In order to more easily understand the effects of spin evolution alone, we do not consider the full model with the self-gravity feedback switch in this discussion (rightmost panels), i.e.~we exclude simulation 5.

Using the fiducial model as a reference, we find that the BH mass -- stellar mass function in our model follows the same trend and is consistent with the observational fit of \citet{Kormendy2013}. A pronounced change in the accretion rate in the BH population at BH masses of about $10^{8}\,{\rm M}_{\odot}$ appears in both simulations as consequence of the fiducial feedback switch (equation \ref{eq:CanonicalTransition}). Given that the radiative efficiency has a strong non-linear dependence on the spin parameter (equation \ref{eq:Epsilon_R}), periods of chaotic accretion can alter the BH mass accretion history significantly. This is evinced in the larger dispersion in the relation for our model. However, we remark that gas accretion is still self-regulated correctly by feedback for BH masses smaller than $10^{8}\,{\rm M}_{\odot}$ since the global slope of the relation is preserved. For BH masses larger than $10^{8}\,{\rm M}_{\odot}$, gas accretion is suppressed and mass growth is driven by mergers with lower mass BHs, which naturally leads to a scaling relation.

For the BH mass function, both simulations exhibit similar trends that are consistent with the observational constraints compiled by \citet{Shankar2009} from X-ray and optical measurements. The only obvious discrepancy with observations is the overabundance of BHs of masses of $10^{7.8} - 10^8 \,{\rm M}_{\odot}$ present in both the fiducial model and our model. In order to check the validity of this finding, we recall the observational galaxy colour -- stellar mass relation and the redshift independent transition from the blue to the red sequence at a stellar mass of about $10^{10}\,{\rm M}_{\odot}$ \citep{Baldry2006, Peng2010}. This stellar mass scale corresponds to a BH mass of $10^8\,{\rm M}_{\odot}$ according to the BH mass -- stellar mass relation \citep{Kormendy2013}, which implies that the activation of an efficient quenching mechanism (e.g. our kinetic feedback mode) must occur precisely at the BH mass scale at which we find the BH overabundance in our simulations. We hypothesise that the origin of this feature is the suppression of gas accretion induced by kinetic feedback at this specific mass scale, which causes a ``bottleneck'' effect in BH growth as BH binary coalescence is abruptly left as the only active mass accretion channel.

We also compute separately the mass function of BHs that have been in the kinetic feedback mode at least once during their lifetime (red histograms) and BHs that have been always in the quasar feedback mode (blue histograms). We obtain again almost an identical result as in the fiducial model. For instance, we find that the red histograms extend over almost the entire BH mass range, which means that the kinetic feedback mode plays always a role in BH growth. However, for low mass BHs, this role is subdominant as this mode only fully kicks in for BHs more massive than $10^8\,{\rm M}_{\odot}$. We return to this result in Section~\ref{SecBHSpinFeedback} as it represents one of the more significant differences when our full model with the self-gravity feedback switch is used.

%.........................................................................
%	Figure 8
%.........................................................................
%BH mass relations
%.........................................................................
 
%--------------------------------------------------------------------------------------------------
\subsection{Galaxy mergers and BH binary coalescence}
%--------------------------------------------------------------------------------------------------

\begin{figure}
\centering
\includegraphics[trim={0cm 0cm 0cm 0cm},width=0.45\textwidth]
{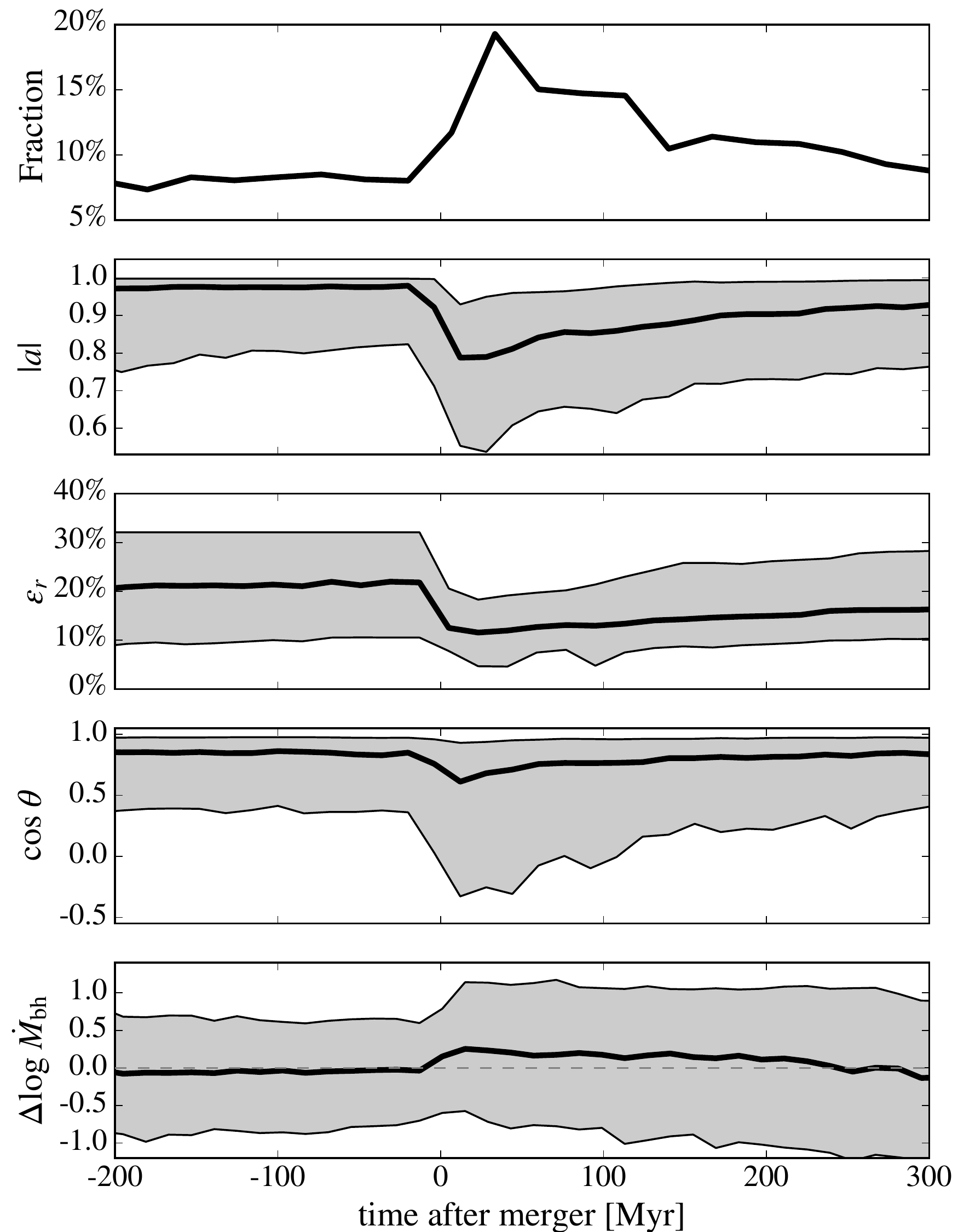}
\caption{\small{Average time evolution of different properties of merging BHs relative to the galaxy coalesence time. Negative time values indicate pre-merger stages, where only the more massive BH in the pair is shown. Also, we restrict our analysis to BHs that are in the coherent accretion regime before the merger. \textit{From top to bottom:} Probability of a counter-rotating accretion episode, stacked median profiles of BH spin, radiative efficiency, angle between BH spin and angular momentum of surrounding gas, and fractional change of the mass accretion rate. For the last property, we normalise each individual profile with the mass accretion rate in the fiducial model $50\, \mathrm{Myr}$ before the merger time. The grey regions are enclosed by the $25\%-75\%$ percentiles of the corresponding distributions.}}

\label{fig:MergerAnalysis}
\end{figure}

In this subsection, we study the effects of our spin model on the population of coalescing BH binaries. We start by analysing the evolution of different BH properties during galaxy mergers. In order to do so, we centre every BH binary history at the BH coalescence time, with negative values indicating pre-merger stages. We finally stack all individual profiles to obtain general trends. For pre-merger stages, we restrict our analysis to the most massive BH of each binary, with the additional constraint of being in the quasar feedback mode before coalescence. This guarantees that gas accretion is still active. We show the results in Figure~\ref{fig:MergerAnalysis}.

We compute the fraction of BHs with counter-rotating accretion in the first panel. This is done by counting how many BHs experience counter-rotating accretion in every time bin. At about $20\, \mathrm{Myr}$ before coalescence, we find that the fraction monotonically increases until a maximum of $20\%$, which is reached at about $40\, \mathrm{Myr}$ after coalescence. The fraction drops to pre-merger values after $300\, \mathrm{Myr}$. Counter-rotating accretion becomes more likely to occur due to the rapid fluctuations in the central gas angular momentum induced by gas inflows and turbulence during galaxy mergers \citep{Mayer2007}. This situation was already encountered in the individual BH history analysed in Figure~\ref{fig:HistoryBH1}. In the second, third and fourth panels we plot the median trends for the magnitude of the spin parameter, the radiative efficiency and the angle between the gas angular momentum and the BH spin, respectively. There, we also find evidence of a rapid spin misalignment, i.e.~the spin parameter drops to a median value of about $0.8$, and with it, the radiative efficiency decreases from $20\%$ to about $10\%$. In some cases, galaxy mergers are also able to trigger the self-gravity regime, which by construction leads to incoherent accretion episodes.

In the last panel, we follow the fractional change of the BH mass accretion rate with respect to the fiducal model, $\Delta \log \dot M_{\mathrm{bh}} \equiv \log \dot M_{\mathrm{bh}} - \log \dot M_{\mathrm{bh,fid}}$. To do so, we match every BH binary with its counterpart in the fiducial simulation; then, we follow the merger in both simulations and compare the mass accretion rate histories centred on the respective coalescence times. For pre-coalescence times, we find that BH accretion in our simulations follows the same trend as in the fiducial model as the median radiative efficiency coincides with the fixed fiducial value ($\epsilon_r=0.2$). For post-merger times, the drop in radiative efficiency has a twofold effect on BH accretion. On one hand, less radiated energy means more matter is available to the BH. Besides, a temporarily diminished BH feedback aids the global accretion onto the BH-disc system. On the other hand, some BHs are accreting at Eddington rates as we select only those in the quasar feedback mode before coalescence. The inverse proportionality of Eddington rate on radiative efficiency implies that mass accretion in those BHs is capped with a higher value. All this results in enhancements of up to $0.25\, \mathrm{dex}$ in BH accretion a few tens of Myr after coalescence with respect to the fiducial case.

%.........................................................................
%	Figure 9
%.........................................................................
%Stacked Mergers
%.........................................................................

\begin{figure}
\centering
\includegraphics[trim={0cm 0cm 0cm 0cm},width=0.5\textwidth]
{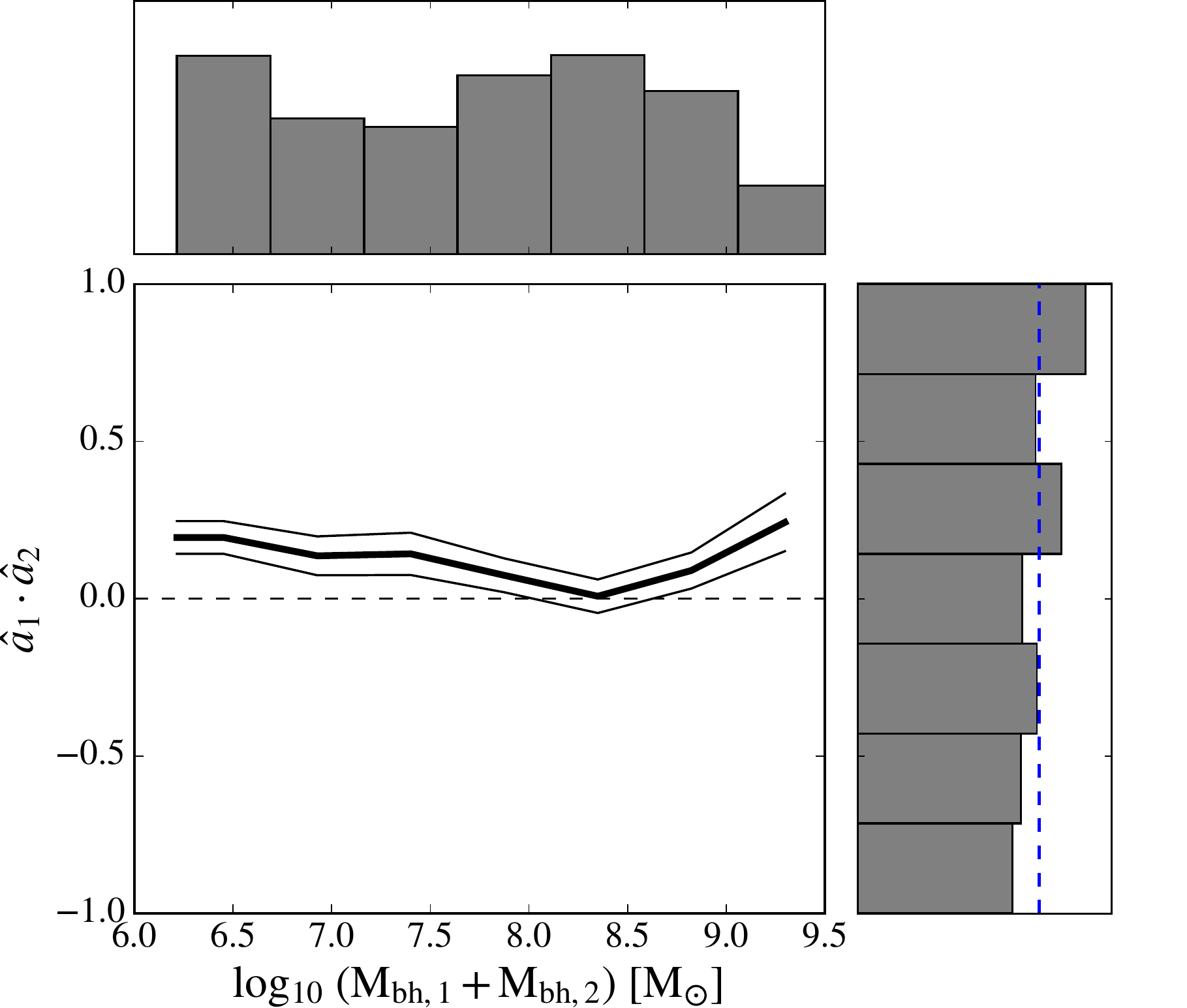}
\caption{\small{\textit{Central panel:} median of the scalar product between the two normalised spin vectors of BH binaries prior to coalescence as a function of their total mass (\textit{solid thick line}) and the standard error of the median (\textit{solid thin lines}). \textit{Top} and \textit{right panels}: 1D histograms of BH binary masses and spin alignment, respectively. The \textit{blue dashed line} represents an isotropic and uncorrelated distribution of the two spins.}}

\label{fig:AlignmentSpinsBinaries}
\end{figure}

We proceed by analysing the alignment of the two BH spins in BH binaries prior to coalescence as a function of binary mass. The results are shown in Figure \ref{fig:AlignmentSpinsBinaries}. For almost the entire mass range, the spins in BH binaries tend to be slightly aligned as the trend goes towards positive values. The only exception appears in BHs of masses $10^8 - 10^{8.5}\, {\rm M}_{\odot}$, in which we find a more isotropic distribution. In this mass range, the self-gravity regime also kicks in, and given that we adopted the chaotic accretion scenario (i.e. $k=0$), the BH spin orientation is randomised. For less massive BH binaries, their host galaxies are mostly gas rich, and therefore, galaxy mergers are frequently wet. This implies that the two BHs in a binary shared a common gas reservoir before coalescence, and through gas accretion, both spins can be orientated in a similar direction. For more massive binaries, gas accretion is already suppressed and most galaxy mergers are dry. Spin alignment in this case is somewhat unexpected and therefore arguably more interesting. We hypothesise that galaxy alignment with the cosmic web might play a role here. For instance, \citet{Mesa2018} find that the axis connecting SDSS galaxy pairs is aligned with the host cosmic filament, showing thus that the galaxy merger process is not isotropic and depends on environment. The positive spin alignment in BH binaries might then be an imprint of a large-scale galaxy alignment.

%.........................................................................
%	Figure 10
%.........................................................................
%Recoil Alignment
%.........................................................................

%.........................................................................
%	Figure 11
%.........................................................................
%Recoil Velocity
\begin{figure}
\centering
\includegraphics[trim={0cm 0cm 0cm 0cm},width=0.5\textwidth]
{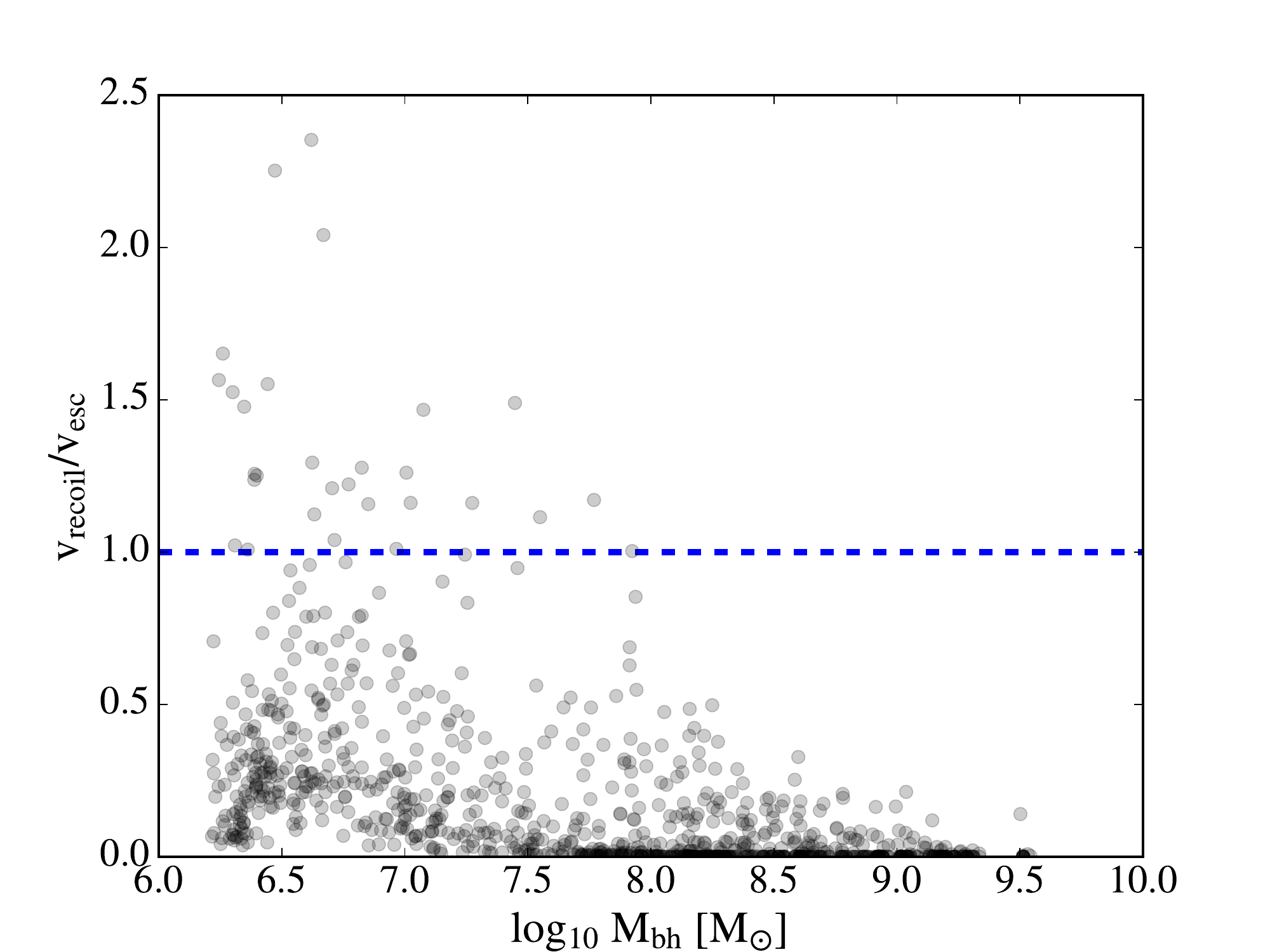}
\caption{\small{Recoil velocities of BH binary remnants normalised by the local escape velocity of the host galaxy as a function of BH mass. The \textit{blue dashed line} represents the threshold above which free floating BHs can potentially form.}}
\label{fig:RecoilVelocity}
\end{figure}
%.........................................................................

Finally, we plot in Figure \ref{fig:RecoilVelocity} the distribution of gravitational recoil velocities for remnants of coalescing BH binaries as a function of BH mass. We normalise with the local escape velocity of each host galaxy as derived from the central gravitational potential; therefore, individual systems with $v_{\mathrm{recoil}}/v_{\mathrm{esc}}>1$ are BHs that can potentially escape their host and form a population of free floating BHs \citep{Favata2004, Gonzales2007b}. This is indeed the case for a few BHs with masses below $10^{8}\, {\rm M}_{\odot}$. For higher mass BHs, their host galaxies are massive enough to keep them gravitationally bound. The recoil produced by a gravitational wave burst depends on the mass asymmetry of the BH binary and the BH spins (see equation~\ref{eqn:vrecoil}). The term depending on the mass asymmetry only accounts for a small fraction of the recoil velocity and is not enough to eject the BH remnant. Consequently, the configuration of the spins in a BH binary is what most likely drives the strong recoil that ultimately produces a free floating BH \citep{Campanelli2007}. Milder recoil kicks, which are more ubiquitous in almost the entire mass range, are also extremely important as they yield a population of off-centred BHs, which can impact lifetimes of AGNs, enhance central star formation, ease the formation of massive stellar cusp, and increase the intrinsic scatter of the $M_{\sub{bh}}-\sigma_*$ relation \citep{Sijacki2011, Blecha2011}. Unfortunately, we cannot assess these effects self-consistently as BHs are artificially glued to the galaxy potential minima in our numerical model and thus the recoil velocity is not applied. Our result therefore merely demonstrates a capability of our model that has yet to be exploited self-consistently.

As a caveat, we also remind the reader that our treatment of BH coalescence is simplistic and does not account for processes such as dynamical friction, stellar hardening and circumbinary discs, which can exert an important influence on BH binaries. For example, \citet{Dotti2010} show that large-scale nuclear discs, which are likely to form in gas-rich mergers, can orientate both BH spins perpendicular to the binary orbital plane in a relatively short time scale ($\lesssim 10\, \mathrm{Myr}$), which ultimately results in weaker gravitational recoil kicks compared to an isotropic spin distribution. The positive alignment of BH binaries in our simulation (Figure~\ref{fig:AlignmentSpinsBinaries}) reinforces this conclusion as it shows that spins are already mildly oriented, even without including a circumnuclear disc. In the light of this discussion, our estimates of the velocity recoil of BH binary remnants in gas rich mergers are likely overestimated, and we consequently interpret them as an upper limit case. A proper treatment in which dynamical friction and other dissipative processes are included along with a self-consistent model of spin evolution is essential to properly address the effects of off-centred and free floating BHs in a cosmological context. We plan to address this in future work.

%==================================================================================================
\section{Black hole spin and feedback}  \label{SecBHSpinFeedback}
%==================================================================================================

\begin{figure*}
\centering
\includegraphics[width=1.0\textwidth]{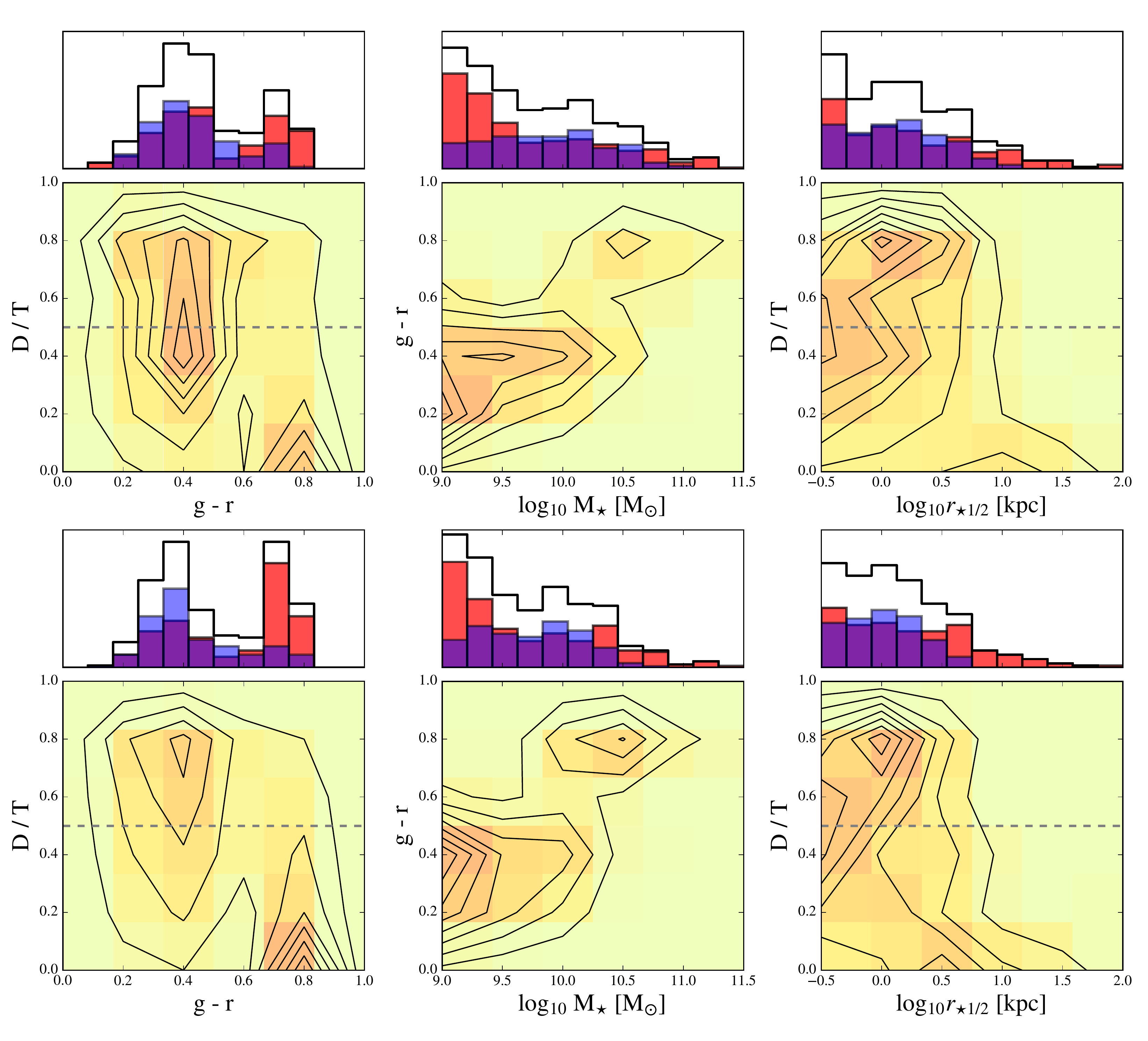}
\caption{\small Morphology -- colour (\textit{right panels}), colour -- stellar mass (\textit{central panels}) and morphology -- size (\textit{left panels}) relations at $z=0$ for the fiducial model (\textit{top panels}) and our full model assuming chaotic accretion in the self-gravity regime (\textit{bottom panels}). For each property, we take 6 bins in the shown ranges which are used to compute the 2D histograms and the contours. In the top 1D histograms of each panel, the blue histograms correspond to galaxies with $\mathrm{D/T} \geq 0.5$, which we vaguely associate with early-type morphologies. The red histograms correspond to galaxies with $\mathrm{D/T} < 0.5$, which exhibit late-type morphologies.}
\label{fig:GalaxyPopulationProperties}
\end{figure*}

In this section, we explore the impact of our novel BH feedback switch, which is based on the self-gravity of the accretion disc, on the galaxy population. To this end, we plot in Figure~\ref{fig:GalaxyPopulationProperties} the morphology -- colour relation (right panels), the colour -- stellar mass relation (central panels), and the morphology -- size relation (left panels), for the fiducial run (top panels) and for our full model with BH spin evolution and the self-gravity feedback switch (bottom panels). We quantify galaxy morphology with the disc-to-total mass ratio (D/T), which is computed from the total mass of stellar particles within one tenth of the virial radius ($0.1\, R_{200,c}$) that have relatively high circularity ($\epsilon>0.7$) and do not belong to the spheroidal component. For galaxy colour we use the $\mathrm{g-r}$ index. Galaxy stellar mass is computed from all star particles within twice the stellar half mass radius ($2\, R_{1/2,\star}$), as in Figure~\ref{fig:BHMassRelation}. Finally, the stellar half mass radius is used as a proxy of galaxy size.

We recall that a strong galaxy colour bimodality consistent with observations is found in IllustrisTNG \citep{Nelson2018}. It is thus not surprising that in our fiducial run, which implements the same IllustrisTNG physics model, we also find a strong bimodality with a population transition at stellar masses of about $10^{10}\,{\rm M}_{\odot}$. Interestingly, our new feedback switch produces an even stronger colour bimodality with a population transition also at $10^{10}\, {\rm M}_{\odot}$. Specifically, our model yields a larger population of red sequence galaxies, but with a somewhat lower stellar mass, i.e.~the average stellar mass of galaxies with $\mathrm{g-r}>0.6$ is $10^{10.3}\, {\rm M}_{\odot}$ and $10^{10.6}\, {\rm M}_{\odot}$ in our model and in the fiducial mode, respectively. Galaxies also exhibit smaller sizes in our model.

An analogous result appears for the BH mass -- stellar mass relation and the BH mass function in our full model (see right panels of Figure~\ref{fig:BHMassRelation}), in which BHs more massive than $10^8\, {\rm M}_{\odot}$ seem to have a somewhat stagnant growth. The ``bottleneck'' effect observed in the fiducial model and the model with BH spin and fiducial feedback switch is also present in the full model, and is even stronger. The fact that the BH mass -- stellar mass relation is still followed demonstrates that the underlying mechanism responsible for this effect is also keeping host galaxies from growing their stellar content. Looking at the Eddington ratio vs BH mass diagram (Figure~\ref{fig:DiagramEvol}), we can deduce that a BH of mass $10^8\, {\rm M}_{\odot}$ and accretion rate $\dot M = 0.1\, \dot M_{\sub{Edd}}$, which is a common scenario at high redshift ($z\sim 2$), would be promptly put into the kinetic feedback mode in our model, whereas it would remain in the quasar mode if the fiducial switch was adopted instead. This means that stellar quenching and suppression of gas accretion by AGN feedback in massive galaxies are maintained for a longer time in our model.

As described in Section~\ref{SecNumerics}, we adopt the same set of numerical parameters used in IllustrisTNG, which includes the coupling efficiency of the kinetic feedback mode ($\epsilon_{\sub{f, kin}} = 0.2$). Because this parameter quantifies how much energy is ultimately delivered to the gas in the form of kinetic energy, its specific value can in principle be set by tuning the simulation to reproduce specific observational trends related to BH-galaxy co-evolution. Given that we inject kinetic feedback energy to the gas at the same rate, but for more prolonged time periods than the fiducial model, we are likely overdoing AGN galaxy quenching in massive galaxies, which results in an atrophied BH and galaxy population. A potential solution to this problem is to scale down the value of $\epsilon_{\sub{f, kin}}$ in order to deliver a similar amount of total kinetic energy as in the fiducial model, but in a more steady and prolonged fashion. A second, and perhaps more realistic alternative is to consider a spin dependent kinetic feedback (see equation~\ref{eq:JetPower}). The feedback energy will be initially very powerful as a BH just entering the self-gravity regime should have a high spin value from the coherent regime. Nevertheless, once stochastic accretion in the self-gravity regime sets in, the BH will be rapidly spun down, and thus the feedback energy will diminish in a few hundreds of Myr. In this way, we could potentially produce a more consistent galaxy population at $z=0$, while at the same time account for high redshift galaxy quenching.

Although no parameter re-calibration was carried out in our simulations for the purpose of this initial study, our results are in remarkably good agreement with the fiducial model. An interesting discrepancy arises in the morphology -- colour relation, however. As discussed in Section~\ref{SecBHmodel}, galaxies in IllustrisTNG exhibit morphologies inconsistent with their colours, i.e. there is an excess of red discs and blue spheroids \citep{Rodriguez-Gomez2018}. This is indeed seen in our fiducial simulation, too, in which 
bluer galaxies ($\mathrm{g-r}<0.6$) do not exhibit a clear morphological trend, i.e.~there is an almost equal fraction of spheroids and disc-like galaxies that are blue. In our model, however, a clearer trend towards disc-like morphologies appears for blue galaxies, which is more consistent with observations. The population of red galaxies is also clearly dominated by spheroids. This result is very interesting because, in spite of the somewhat stronger kinetic feedback in our simulation, our self-gravity feedback switch still manages to select the correct morphological type of galaxies to quench, even if done slightly too strongly.

%.........................................................................
%	Figure 12
%.........................................................................
%Galaxy properties
%.........................................................................

\begin{figure}
\centering
\includegraphics[trim={0cm 0cm 0cm 0cm},width=0.5\textwidth]
{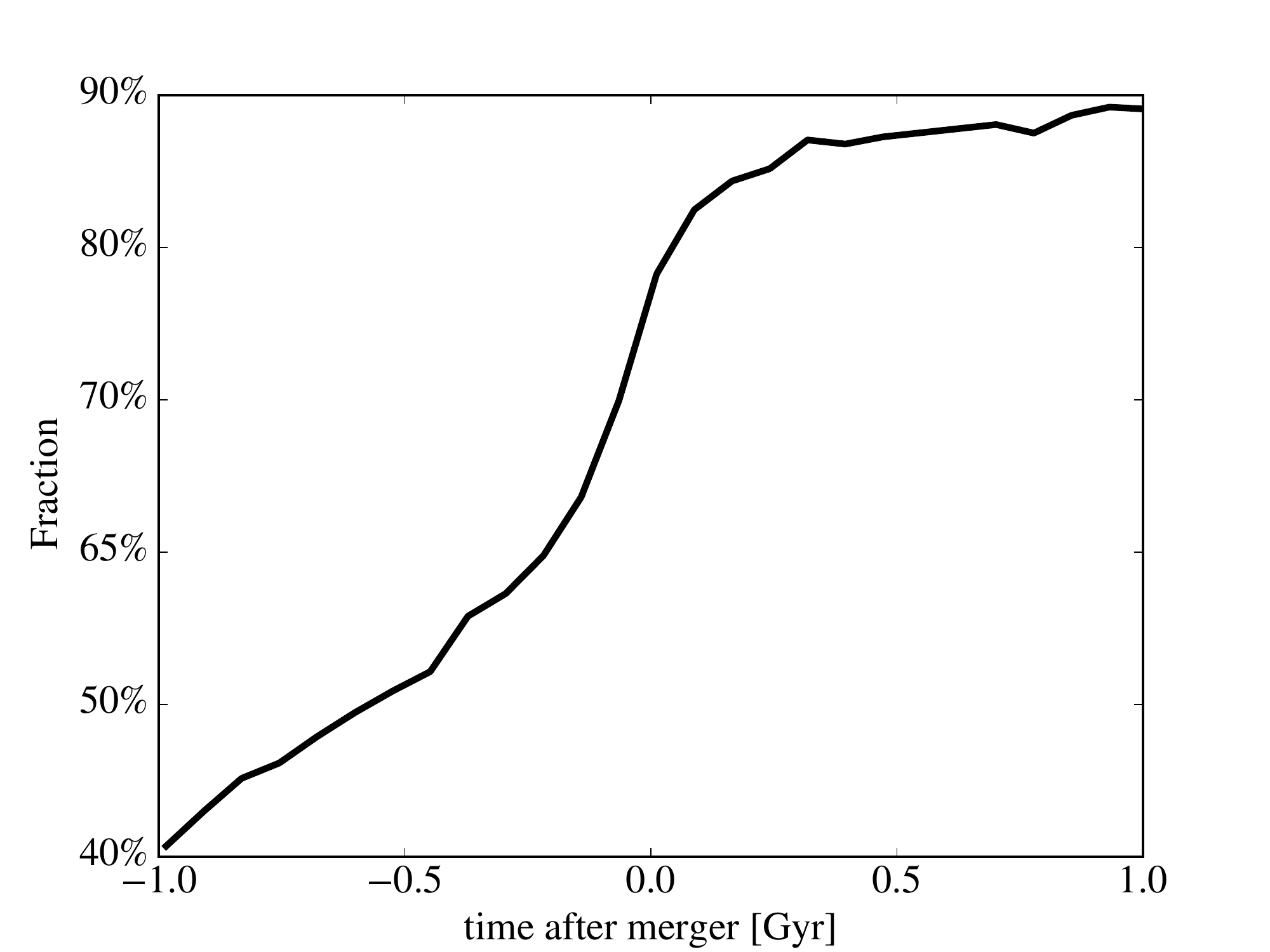}
\caption{\small{Fraction of BHs of masses $10^{7.5} - 10^{8.5}\,{\rm M}_{\odot}$ hosted by  merging galaxies that are in the self-gravity regime as a function of time. Negative time values indicate pre-merger stages, where the merger time is defined as the coalescence of the two involved central BHs. Note that only mergers of galaxies massive enough to host BHs in the selected mass range are considered.}}

\label{fig:FractionSG_Mergers}
\end{figure}

With the aim of further investigating the previous result, we compute in Figure~\ref{fig:FractionSG_Mergers} the evolution of the fraction of BHs in the self-gravity regime that have masses between $10^{7.5}\, {\rm M}_{\odot}$ and $10^{8.5}\, {\rm M}_{\odot}$ and are hosted by merging galaxies. This specific mass range is where most BHs enter the self-gravity regime and, therefore, hosts galaxies which become quenched and red. We notice a significant increase in the BH fraction from $500\, \mathrm{Myr}$ before to $500\, \mathrm{Myr}$ after the merger time. This demonstrates that galaxy mergers speed up the onset of the self-gravity regime, because a boosted BH growth, either through gas accretion or BH coalescence or both, facilitates the transition from the coherent regime (see Figure~\ref{fig:DiagramEvol}). For our model, this has the consequence that the kinetic feedback mode is activated predominantly in merging galaxies, in which major morphological changes take place as well. This explains why we recover the galaxy morphology -- colour relation in our simulations, while still reproducing the colour bimodality and the BH mass -- stellar mass relation.

%.........................................................................
%	Figure 13
%.........................................................................
%Fraction of BH in self-gravity
%.........................................................................

%==================================================================================================
\section{Summary and Conclusions}  \label{SecConclusions}
%==================================================================================================

We have implemented a new sub-grid model for BH spin evolution in the $N$-body magneto-hydrodynamical moving-mesh code {\small AREPO}. We account for two different channels of spin evolution, namely gas accretion and BH binary coalescence. Additionally, we test a hypothesis in which the self-gravity of the BH accretion disc regulates the transition from a quasar BH feedback mode to a kinetic BH feedback mode. We have run several cosmological simulations to explore the new models, including a fiducial run with the original IllustrisTNG physics that is used as a comparison reference. Our main findings can be summarized as follows:

\begin{itemize}
 \item Gas accretion in low and intermediate mass BHs ($M_{\sub{bh}}\lesssim 10^8\, {\rm M}_{\odot}$) occurs in a coherent fashion as small-scale turbulence induced by a self-gravitating disc is negligible. For higher redshift ($z\gtrsim 2$), the angular momentum of the central gas reservoir around the BH can fluctuate considerably. However, low-mass BHs can be easily reoriented, and thus, although the BH spin direction can fluctuate, the net effect is an increase of the spin magnitude. An exception occurs in galaxy mergers, in which strong gas inflows can cause rapid fluctuations in the gas angular momentum that cannot be followed up by the BH, even if it is accreting already at its Eddington rate. This results in a decrease of both the spin parameter and the radiative efficiency, because counter-rotating accretion becomes more likely. A consequence of this is a boost in BH growth of up to $0.25\, \mathrm{dex}$ during the merger phase. A few hundreds of Myr after the merger, accretion becomes coherent again, and the BH is quickly spun up. In some intermediate mass BHs, a galaxy merger can trigger the self-gravity regime due to the boosted BH growth.
 
 \item More massive BHs ($M_{\sub{bh}}\gtrsim 10^8\, {\rm M}_{\odot}$) can develop massive accretion discs in which self-gravity effects become important and counter-rotating accretion occurs frequently. This leads to a drop of the BH spin magnitude that depends on the degree of anisotropy of the incoming directions of accretion episodes that replenish the inner disc. Nevertheless, these BHs usually reside in massive, early-type galaxies in which star formation activity must be promptly quenched. Therefore, most of them are in the kinetic BH feedback mode, and thus, gas accretion is drastically suppressed. This implies that spin evolution is mainly driven by BH binary coalescence, while gas accretion in the self-gravity regime plays only a weak role in shaping the BH spin evolution. Our predicted BH spin distribution is in very good agreement with the observational data compiled by \citet{Reynolds2013}, and with previous studies of BH spin evolution in a cosmological context \citep[e.g.][]{Dotti2013, Dubois2014c}.
 
 \item Due to the non-linear dependence of the radiative efficiency on the BH spin, even short periods of incoherent accretion can significantly impact BH growth. This is seen in the larger scatter of the BH mass -- stellar mass relation in our simulation compared to the fiducial TNG model, in which a calibrated fixed value for the radiative efficiency is adopted. The fact that our relation still follows the same trend as the fiducial model, and is consistent with the observational fit of \citet{Kormendy2013}, reveals however that self-regulated gas accretion still operates in a broadly similar fashion if BH spin evolution is accounted for.
 
 \item The spins of BH binaries are mildly aligned prior to coalescence, with the only exception occurring in BH binaries of masses $10^8\, {\rm M}_{\odot}$ to $10^{8.5}\, {\rm M}_{\odot}$, which exhibit a more isotropic distribution due to incoherent accretion induced by the disc self-gravity. For low-mass BH binaries, their coalescence occurs usually in gas-rich environments, in which a common gas reservoir can orient, through gas accretion, both spins in a similar direction. Further alignment can be reached by the action of a circumnuclear disc \citep{Dotti2010}. For more massive BHs, gas accretion is already suppressed by the strong BH feedback. Therefore, spin alignment in this case might presumably reflect a large-scale alignment of galaxies with their environment. An alignment of BH spins can weaken gravitational wave bursts from BH binary coalescence and the subsequent recoil kick of the BH remnant, and has thus potentially direct consequences for the population of off-centred and free floating BHs, as well as for the predicted gravitational wave background.
 
 \item We propose a novel scheme for activating the kinetic BH feedback mode that is based on the self-gravity of the accretion disc. In this scenario, massive accretion discs, in which self-gravity becomes significant, get fragmented outside their self-gravity radius $R_{\sub{sf}}$. As the disc gets more massive, the self-gravity radius shrinks, and so does the radiatively efficient innermost disc. This leads to a natural shut down of the quasar feedback mode. On the other hand, incoherent accretion induced by disc fragmentation provides favourable conditions for a spherical hot accretion flow. Furthermore, counter-rotating accretion becomes more frequent, in which case the production of non-thermal jet outflows is highly efficient \citep{Garofalo2010}. Our conjecture is therefore to activate the kinetic feedback mode once $R_{\sub{sg}}\leq R_{\sub{warp}}$. We compare with the warp radius $R_{\sub{warp}}$ as only the material inside the warped disc can effectively transfer angular momentum to the BH \citep{Volonteri2007}, i.e.~self-gravity does not significantly impact mass and angular momentum accretion if $R_{\sub{sg}} > R_{\sub{warp}}$. For a highly spinning BH ($a=0.998$), the feedback transition occurs at a BH mass of about $10^8\, {\rm M}_{\odot}$, which, according to the BH mass -- stellar mass relation \citep{Kormendy2013}, corresponds to a host galaxy of stellar mass $\sim 10^{10}\, {\rm M}_{\odot}$. This stellar mass coincides surprisingly well with the mass scale at which galaxies enter the red sequence \citep{Peng2010}, thereby making our proposed scenario a plausible explanation for the onset of star formation quenching in massive galaxies.
 
 \item In order to test our new feedback switch, we analyse different properties of the galaxy population and compare them with the fiducial TNG model. We find a bimodality in the galaxy colour relation that is consistent with the fiducial model, although being somewhat more pronounced. Our high mass galaxies are in general less massive and have smaller sizes, which indicates a stronger kinetic feedback in our simulations. Our proposed feedback switch activates the kinetic feedback mode in massive BHs earlier than in the fiducial model. Considering that we adopted the same set of parameters as in IllustrisTNG, including the coupling efficiency of the kinetic feedback mode $\epsilon_{\sub{f, kin}}$, we are consequently delivering more kinetic energy to the central gas. A possible solution to this problem would be to scale down the efficiency, either by simply adopting a lower value, or by implementing a spin-dependent kinetic feedback (e.g. equation~\ref{eq:JetPower}), in such a way that the kinetic feedback energy is injected to the gas in a more prolonged fashion.
 
 \item \citet{Rodriguez-Gomez2018}  reported a tension between IllustrisTNG and observations with respect to the galaxy morphology -- colour relation,  with red discs and blue spheroids appearing to be in excess in the simulations. As expected, our fiducial run displays a similar trend, as it is based on the original IllustrisTNG physics model. When we use our new model with BH spin evolution and the kinetic feedback switch, galaxies exhibit morphologies more consistent with their colours, i.e.~blue galaxies are predominantly disc types, while red galaxies are mostly spheroids. We find that galaxy mergers are efficient triggers of self-gravity in the accretion discs as a boosted BH growth facilitates the transition from the coherent accretion regime (see Figure~\ref{fig:DiagramEvol}). Considering that we activate the kinetic feedback mode at the onset of the self-gravity regime, and that merging galaxies undergo significant morphological changes, this explains why our approach succeeds in recovering the missing relation between morphology and colour.
 
\end{itemize}

Our implementation of a self-consistent BH spin evolution model into realistic hydrodynamical cosmological simulations is an important advance with respect to the first generation of BH models in which BHs were simply treated as massive sink particles, ignoring the spin. Our results demonstrate the importance of coupling the spin to the physics of BH accretion and feedback, especially during galaxy mergers. The potential role of the BH spin, through the self-gravity of the accretion disc, in shaping the galaxy colour bimodality and a correct morphology -- colour relation is a further tantalising finding, suggesting that future  simulation campaigns should include refined models of BH spin evolution in order to yield more realistic results.

%==================================================================================================
\section*{Acknowledgements}
%==================================================================================================

SB acknowledges support from the International Max-Planck Research School for Astronomy and Cosmic 
Physics of Heidelberg (IMPRS-HD) and financial support from the Deutscher Akademischer 
Austauschdienst (DAAD) through the program Research Grants - Doctoral Programmes in Germany 
(57129429). VS acknowledges the European Research Council through ERC-StG grant EXAGAL-308037, and 
the SFB-881 `The Milky Way System' of the German Science Foundation. Authors like to thank the Klaus 
Tschira Foundation. We have used NASAs ADS Bibliographic Services. Most of the computations that 
made possible this work were performed with {\tt Python 2.7} and their related tools and libraries, 
{\tt iPython} \citep{Perez-2007}, {\tt Matplotlib} \citep{Hunter-2007}, {\tt scipy} and {\tt numpy} 
\citep{Van-2011}.

\bibliographystyle{mn2e}

\label{lastpage}

\end{document}